\documentclass[fleqn,usenatbib]{mnras}

\usepackage[T1]{fontenc}
\usepackage{ae,aecompl}

\usepackage{graphicx}	
\usepackage{amsmath}	
\usepackage{amssymb}	
\usepackage{multirow}

\title[Modelling time-averaged and lag-energy spectra in AGN]{Relativistic X-ray reverberation modelling of the combined time-averaged and lag-energy spectra in AGN}

\author[P. Chainakun, A. J. Young \& E. Kara]{P. Chainakun$^1$\thanks{E-mail: \href{mailto:phxpc@bristol.ac.uk}{phxpc@bristol.ac.uk}}, A. J. Young$^1$ \& E. Kara$^2$\\
$^1$H. H. Wills Physics Laboratory, Tyndall Avenue, Bristol BS8 1TL, UK\\
$^2$Department of Astronomy, University of Maryland, College Park, MD 20742-2421, USA}

\date{Accepted XXX. Received YYY; in original form ZZZ}

\pubyear{2016}

\begin{document}
\label{firstpage}
\pagerange{\pageref{firstpage}--\pageref{lastpage}}
\maketitle

\begin{abstract}
General relativistic ray tracing simulations of the time-averaged spectrum and energy-dependent time delays in AGN are presented. We model the lamp-post geometry in which the accreting gas is illuminated by an X-ray source located on the rotation axis of the black hole. The spectroscopic features imprinted in the reflection component are modelled using {\sc reflionx}. The associated time delays after the direct continuum, known as reverberation lags, are computed including the full effects of dilution and ionization gradients on the disc. We perform, for the first time, simultaneous fitting of the time-averaged and lag-energy spectra in three AGN: Mrk 335, IRAS 13224-3809 and Ark 564 observed with \emph{XMM-Newton}. The best fitting source height and central mass of each AGN partly agree with those previously reported. We find that including the ionization gradient in the model naturally explains lag-energy observations in which the 3~keV and 7--10~keV bands precede other bands. To obtain the clear 3~keV and 7--10~keV dips in the lag-energy profile, the model requires either a source height $>5 r_g$, or a disc that is highly ionized at small radii and is colder further out. We also show that fitting the lag or the mean spectra alone can lead to different results and interpretations. This is therefore important to combine the spectral and timing data in order to find the plausible but self-consistent fits which is achievable with our model.    

\end{abstract}



\section{Introduction}

In the standard paradigm, Active Galactic Nuclei (AGN) consist of a central super-massive black hole surrounded by an accretion disc. During the accretion process, gravitational energy is converted to radiations in form of optical and ultra-violet photons. These photons gain more energy after inverse Compton scattering with the relativistic electrons in a corona so their energy peaks at X-ray wavelengths. The X-rays may originate from the base of a highly-energetic jet, which can be modelled in a simplistic way as a point source continuum above the accretion disc \citep[e.g.,][]{Matt1991,Miniutti2004}. Some of the X-ray photons escape to an observer directly and form a ``direct'' continuum spectrum. Other photons illuminated the disc and are reprocessed before being observed as a ``reflection'' spectrum. The most prominent observable features of the X-ray reflection are the Fe K$\alpha$ line at a rest frame energy of $\approx6.4$ keV, the soft excess at $\approx$ 0.3--2~keV, and the Compton hump at $\approx$ 10--30~keV \citep{George1991,Ross1999, Ross2005}. Measuring the spectroscopic features and how the relativistic effects distort these components allow us to probe the emission region closest to the event horizon \citep[see, e.g.,][for a review]{Reynolds2003}.      

The other approach is to assess the rapid variability that relates to the light-travel time delays between variations in the reflection and continuum components. This technique, referred to as the X-ray reverberation, is a powerful tool to map the location and size of the X-ray source \citep{Uttley2014}. If X-rays in AGN illuminate the inner disc and are reprocessed, we expect energy bands dominated by the reflection to be delayed with respect to the bands dominated by the direct continuum. In other words, the soft excess, Fe~K and the Compton hump bands should lag behind the continuum band. The first hint of X-ray reverberation signatures was found in Ark 564 by \cite{Mchardy2007}. The robust confirmation of reverberation was reported by \cite{Fabian2009} in discovery of the soft lags, or the lags of the soft excess band behind the continuum band by $\approx 30$~s in 1H0707-495. Moreover, the first detection of reverberation between the Fe K and continuum band was reported by \cite{Zoghbi2012} in NGC 4151. The number of AGN that exhibits soft excess lags \citep[e.g.,][]{Demarco2013} and Fe K lags \citep[e.g.,][]{Kara2013a, Kara2014} is growing significantly. The reverberation framework is further supported by \cite{Zoghbi2014} who discovered that the Compton hump lags behind the primary continuum in MGC-5-23-16.

Models that predict these lags are still being developed. The construction of realistic theoretical models is based on ray tracing simulations \citep[e.g.,][]{Fanton1997, Reynolds1999, Ruszkowski2000, Dovciak2004b} aiming to compute the delays between the direct and the reflection photons that arrive at the observer. In \cite{Chainakun2012}, we investigated a realistic lamp-post scheme for 1H0707-495 in which the X-ray source was moved along the symmetry axis and computed the frequency-dependent time lags between the direct and reflection spectra. Although the effects of ionization gradients in the disc were taken into account, we found that a more complex source geometry is required in order to understand its variability. \cite{Wilkins2013} applied the concept of a response function to compute the frequency-dependent time lags for different source geometries. They deduced that in case of 1H0707-495 the X-ray source has a radial extent of $\approx 35r_g$. The first systematic fitting of the lag-frequency spectra under the lamp-post geometry has been performed by \cite{Emmanoulopoulos2014} for 12 AGN in total. They found that, fitting the lag-frequency spectrum alone, the average source height in those AGN is quite small, $\approx 4r_{g}$. \cite{Cackett2014} were the first to fit the energy-dependent time lags of AGN. They computed the response of the Fe K photons and focused on the lags of the Fe K band in NGC 4151. Recently, fitting the combined spectroscopic and timing data of AGN has been performed, for the first time, by \cite{Chainakun2015} in Mrk~335 during its high flux state. Simultaneously fitting the energy integrated spectrum and the lag-frequency spectra both in soft (0.3--1 vs. 1--4 keV) and hard (2.5--4 vs. 4--5.6 kev) bands revealed the source height is small, consistent with the findings of \cite{Emmanoulopoulos2014}.      

In this paper we present a self-consistent model to simultaneously fit the averaged and lag-energy spectra of three AGN, Mrk~335, IRAS~13224-3809 and Ark~564. We trace photon paths along the Kerr geodesics, and use the entire X-ray spectrum so that the full-dilution effects can be included. We also consider the effects of ionization gradients in the disc which play an important role in the spectroscopic and timing profile. The theoretical lag-energy spectrum is produced using a similar technique to that used on the observational data (i.e., compute the lags of each energy bin comparing to the entire band where the energy bin of interest is excluded, as discussed in \citealt{Zoghbi2011, Zoghbi2013}). For Mrk~335 ($z= 0.0258$), we focus on its high flux state observed by \emph{XMM-Newton} in 2006 \citep{Kara2013a}. Building on previous results, such as \cite{Emmanoulopoulos2014, Chainakun2015}, in this paper we fit the time-averaged spectrum simultaneously with the lag-energy spectrum. The spectrum of Mrk~335 has shown not only inner disc reflection but also distant reflection from a cold torus and ionized gas filling that torus \citep[e.g.,][]{Oneill2007, Chainakun2015}. For IRAS 13224-3809 ($z= 0.066$), the Fe K lags have been detected at the frequencies 5.8--10.5 $\times 10^{-4}$ Hz using 500 ks light curves observed by \emph{XMM-Newton} in 2011 \citep{Kara2013b}. \cite{Fabian2013} investigated the corresponding spectrum of IRAS 13224-3809 and found that it could be explained by a patchy disc model. More recently, \cite{Chiang2015} performed a series of spectral fitting of of IRAS 13224-3809 and deduce the black hole mass to be $\approx 3.5 \times 10^6 M_{\odot}$. For Ark 564 ($z=0.0247$), the Fe K lags have been found at frequencies 3.2--5.2 $\times 10^{-4}$ Hz using eight \emph{XMM-Newton} observations in 2011 \citep{Kara2013a}. \cite{Giustini2015} investigated the energy-integrated spectra of these eight \emph{XMM-Newton} observations but also include one Suzaku observation. They found the averaged spectrum is complex. Although reflection and absorption-dominated models can reproduce time-averaged data, both require contrived geometries (e.g., the warm absorber covers only the distant reflection component). Combining the averaged and lag-energy spectral fitting has not been performed before for AGN, which will be done in this work.

The rest of this paper is organized as follows. The observational data used here for Mrk 335, IRAS 13224-3809 and Ark 564 are summarized in Section 2. We present a theoretical reverberation model and discuss how the model predicts the lag-energy spectrum in Section 3. The fitting procedure is explained in Section 4 followed by the results in Section 5. The discussion and conclusion are drawn in Sections 6 and 7, respectively.

\section{Observations}

The X-ray data for these three AGN are from the \emph{XMM-Newton} observatory \citep{Jansen2001}. The light curves used in this work and data reduction are similar to those have been analysed in previous literature. For Mrk 335, we use the 133 ks archival data on 2006 January 03, leaving 120 ks of good data after screening the background flare \citep{Kara2013a}. For IRAS 13224-3809, the 500 ks light curves observed over four orbits from 2011 July 19 to 2011 July 29 are used, having 300~ks of clean data \citep{Kara2013b}. We consider the light curves of Ark 564 from eight archival data observed between 2011 May 24 and 2011 July 01 as presented in \cite{Kara2013a}. 

\section{Theoretical modelling of the X-ray reverberation}

We consider the lamp-post geometry \citep[e.g.,][]{Matt1991} in which an isotropic point source is stationary on the symmetry axis of a black hole. The gravitational units of time, $t_g = GM/c^3$, and distance, $r_g = GM/c^2$, are used where $G$ is the gravitational constant and $c$ is the constant speed of light. To minimize the free parameters, the black hole spin is fixed at the physical maximum value, $a = 0.998$. The accretion disc is assumed to be a standard geometrically thin, optically thick disc \citep{Shakura1973} in prograde Keplerian orbit whose radial extent ranges between the innermost stable circular orbit, $r_\text{ms}$, and $400r_g$. A stationary observer is set at a distance $1,000 r_g$ from the central black hole. The source produces the primary X-ray continuum, traditionally modelled as a power-law, that can either be observed directly or as a ``reflection'' spectrum after it has been back-scattered off the disc. The direct and reflection spectra including other components, if required, such as distant reflection, warm absorbers and black body emission, form the time-averaged spectrum. The time delays between changes in the direct continuum flux and the associated echo from the disc are referred to as the reverberation lags. This Section outlines how we numerically compute the time-averaged spectrum and reverberation lags.

\subsection{Time-averaged spectrum}

We assume the primary continuum is a power-law with a cut-off energy at 300~keV, $F(E) \propto E^{-\Gamma} \exp (-E/300\text{ keV})$. Following the methods described in \cite{Ruszkowski2000, Dovciak2004, Chainakun2012, Chainakun2015}, we perform photon path integrals in parallel on Graphic Processing Units (GPUs)\footnote{NVIDIA K20 Graphic Processing Unit (GPU) cards on the BlueCrystal Supercomputer at the University of Bristol, \url{http://www.bris.ac.uk/acrc/}.}  to trace the photons that travel along the Kerr geodesics between the source, the disc, and the observer. If the photons terminate at the event horizon, they are discarded. The direct spectrum is obtained from those photons that arrive at the observer without scattering off the disc. We model ionization gradients on disc by dividing the disc into small $dr d\phi$ bins in which we separately compute the reprocessing of incident photons. The radial disc density is in the form of a power-law, $n(r) \propto r^{-p}$, assuming $p$ is the disc density index ($p=0$ is the case of a constant density disc). We calculate the total incident flux per unit area of each bin, $F_t(r,\phi)$, so that the corresponding ionization parameter, $\xi = 4 \pi F_t(r,\phi) / n(r)$, is obtained. The {\sc reflionx} model \citep{George1991, Ross1999, Ross2005} is applied to deal with the X-ray reprocessing by the disc. We then have the emergent spectrum from each disc element which is transferred to the observer's frame via the backward ray-tracing technique \citep{Fanton1997}. The redshifts and full-relativistic effects \citep[e.g.,][]{Cunningham1975} are included in our calculations. The total spectrum is the sum of direct and reflection components integrated over all disc bins. The model parameters consist of the source height ($h$), disc inclination angle ($i$), photon index ($\Gamma$), iron abundance ($A$), ionization state at the $r_\text{ms}$ ($\xi_\text{ms}$), density disc index ($p$), black hole mass ($M$) and scale-factor ($R_s$) defined as the ratio between the reflection flux and direct continuum flux measured in 5--7~keV band. The direct and reflection spectra will be normalized to match up with the flux ratio in 5--7~keV band determined by $R_{s}$. In doing so it will systematically reveal the ratio of reflection and continuum flux across all energy bands, and hence in our model there is no preferable band to be selected for the variable parameter $R_{s}$. The variations of the X-ray source produces the variations of direct power-law and reflection components which are assumed to vary coherently. This will be discussed further in the following Section.

\subsection{Energy-dependent time lags}

In the observations, the lag-energy spectrum is extracted in the particular frequency range where the soft lag is found. The lag-energy spectrum shows the relative time lags of all energy bins compared to a reference band. The choice of the reference band determines the absolute time lags, the overall phase of which is arbitrary, so we are instead focusing on relative lags \citep[e.g.,][]{Zoghbi2011,Zoghbi2013}. We compute the model lag-energy spectrum in a similar way to the observational data, i.e., using the entire band as a reference with the energy bin of interest excluded. The band with larger lags is delayed with respect to the band with smaller lags, and the lag equals their relative difference.

We define the Reflected Response Fraction (RRF), similar to \cite{Cackett2014}, as $(\text{reflection flux}) / (\text{continuum flux})$. We define a model parameter $R_s$ that is the RRF measured in the $5-7 \text{ keV}$ band that, once selected, gives the corresponding RRF in each energy band, $R(E)$. Assuming the primary continuum has a normalization that varies in time as $x(t)$, we use the equations below to produce the light curves of a specific energy band of interest (``int'') and a reference band (``ref''), respectively.
\begin{align}
\text{int}(E, t) = F(E) \left[ x(t) + R(E) x(t^\prime) \otimes \psi(E,t^\prime) \right], \label{eq:int} \\ 
\text{ref}(E, t) = \sum_{E_i \ne E} \text{int}(E_i,t), \label{eq:ref}
\end{align} 
where $F(E)$ is the power-law flux and $\psi(E,t)$ is the response function of the energy bin $E$. The convolution is defined as $x(t') \otimes \psi(E,t') = \int_0^t x(t^\prime) \psi(E,t-t^\prime)dt^\prime$. On the right hand side of equation (1), the $1^\text{st}$ term represents the contribution from the direct power-law component which changes directly with the primary variations. The $2^\text{nd}$ term relates to the response from the disc. $R(E)$ is the ratio $(\text{reflection flux}) / (\text{continuum flux})$ in the bin $E$ while $F(E)$ weights the flux across all energy bins. The reference band, equation (2), is the integrated flux across all bins subtracting the flux of the energy bin of interest. The time lags are computed using the standard Fourier technique \citep{No99}. We calculate the cross-spectrum, $C(f) = \text{INT}^{*}(E,f) \text{REF}(E,f)$, where $\text{INT}(E,f)$ and $\text{REF}(E,f)$ are the Fourier forms of the $\text{int}(E,t)$ and $\text{ref}(E,t)$, respectively. The symbol $^*$ denotes the complex conjugate of that component. The time lags are $\tau(E,f) = \arg C(E,f)/(2\pi f)$ where $\arg C(E,f)$ is the argument of the cross-spectrum of the energy bin $E$.  

Note that the lag-energy profile was modelled by \cite{Cackett2014} but they considered only the lags produced by the Fe K$\alpha$ photons from the neutral reflection. Our new approach is taking into account photons of all X-ray energies and considers the effects of ionization gradients in the disc. We include the both direct and reflection flux components in each energy bin, regardless of which component dominates in that bin. In other words, the full effects of dilution are taken into account. Our model is therefore a realistic modification and is calculated in a similar way to the lag-energy observations, and our model is suitable to fit to the data.  

\subsubsection{Testing the model}

\begin{figure}
\centering  
\includegraphics*[width=70mm]{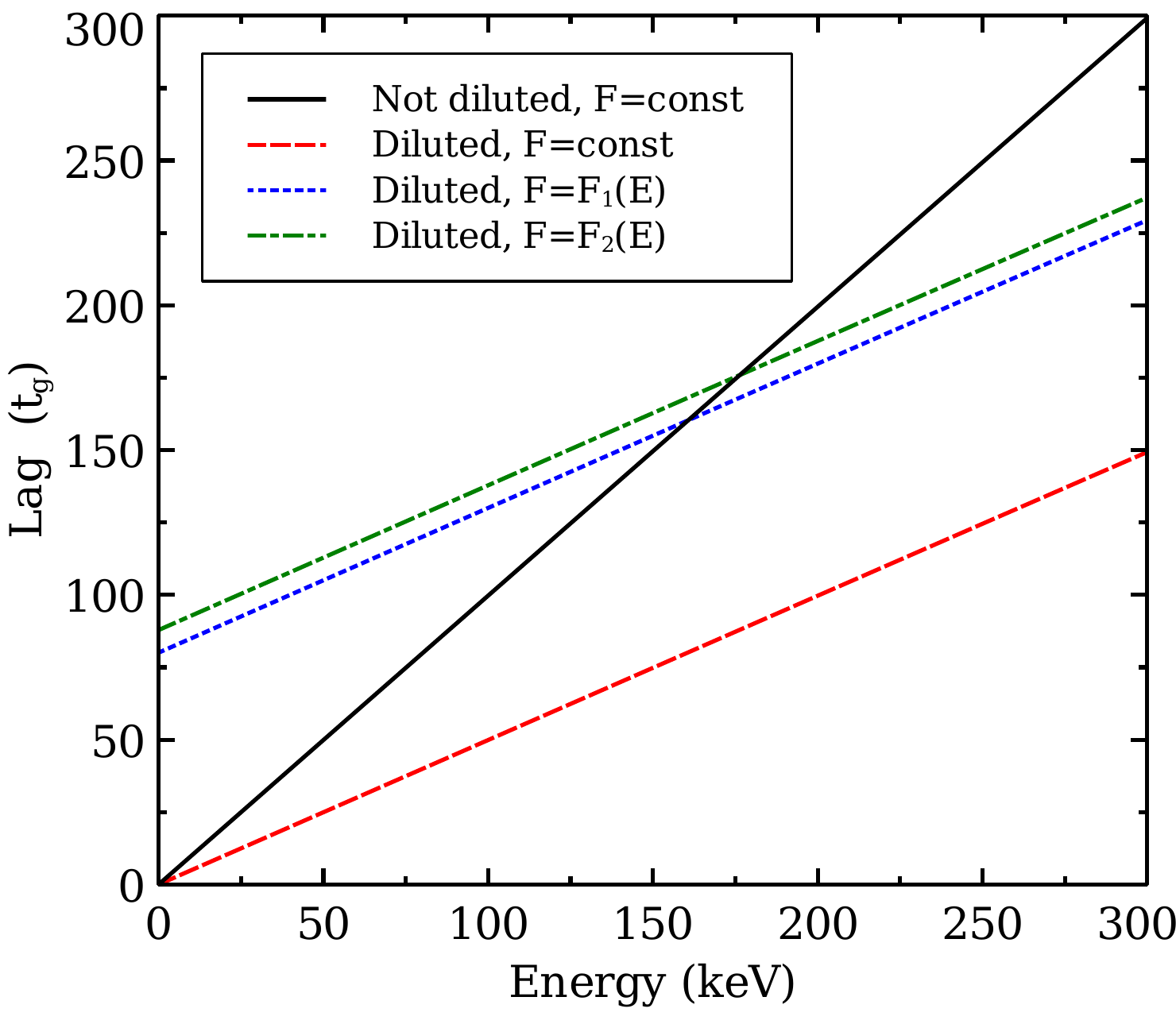}
\caption{Lag-energy plots when the response of each energy bin lags, or leads, other bins by $x\Delta t = \Delta E$ where $x=1$~keV/s is a normalization factor. We first assume a flat continuum, $F = \text{const}$. This is for the purpose of testing the model. In the case of no dilution (solid black line), the lag is a straight line with slope 1, as expected. Once dilution effects are included, the relative lags decrease (red-dashed line). We also show the plots of two different continuum spectra ($F_1(E)$, blue-dotted and $F_2(E)$, green-dashed-dotted lines) whose corresponding lags differ only in overall (rather than relative) normalization. \label{fig:toy-model-1}}
\end{figure}

In order to validate the model and investigate equations (\ref{eq:int}) and (\ref{eq:ref}), let us assume the RRF and F(E) are constant for all energy bands (i.e., $R(E) = R = 1$ and $F(E) = F = 1$). Furthermore, the $1^\text{st}$ term on the right hand side of equation (\ref{eq:int}) is set to be 0 (i.e., no dilution effects contributed by the continuum in the $\text{int}(E,t)$ band). We assume the response function is in the form of a delta function, $\psi(E,t)=\delta(xt-E)$ which equals 1 when $xt = E$ and equals 0 elsewhere. This forces the response in each energy bin to lag, or lead, others by $x\Delta t = \Delta E$ where $x=1$~keV/s is a normalization factor. In this case the lag-energy spectrum should have a linear relation with gradient 1 as the relative lags change mostly with the response function. This is exactly what the model predicts as is shown by the solid black line in Fig.~\ref{fig:toy-model-1}. Including dilution effects could significantly reduce relative lags comparing to their undiluted values. In Fig.~\ref{fig:toy-model-1}, we also show the plots of different continuum models, $F_{1}(E)=\int E^{-1.5}{\rm d}E$ and $F_{2}(E)=\int E^{-2.5}{\rm d}E$. This parameter changes the absolute lags while the relative lags remain the same and, hence, can be neglected since absolute lags in the lag-energy analysis are not relevant. The reason that $F(E)$ does not change the shape of the lags is because the RRF and $\psi(E,t)$ remain the same in all cases. Although, $F(E)$ depends on the photon index, this does not imply that the photon index has no effects on the relative lags. The reflection features change with the photon index so, as will be shown in Fig.~\ref{lags-g}, its effects on relative lags are imprinted in the RRF.      

The dilution effects on time lags have been discussed in \cite{Uttley2014}. We note that the relative lags are absent only if the RRF is constant across all energy bands and $\psi(E,t) = \psi(t)$. If, however, $\psi(E,t)$ is energy dependent relative lags occur regardless of the RRF. Both $\psi(E,t)$ and RRF then play an important role in generating the lag-energy profile. In the following Section we will use the model to investigate the reverberation framework in which $\psi(E,t)$ and the RRF are related to physical geometries and properties of the X-ray source, the disc and the black hole.

\subsubsection{Investigating the reverberation scheme}

The X-ray reverberation in AGN originates through the variability of the X-ray source that results in each disc element responding at a different time. As shown by \cite{Chainakun2015}, the response function from the X-ray reflection scenario is a function of both energy and time, $\psi(E,t)$. We assume the reprocessing by the disc is instantaneous so the energy-dependent reverberation delays mostly depend on the different light-travel time of the direct and reflection photons. The exact, and most realistic, response function to date can be obtained only with ray tracing simulation tracking photon position and time taken along their Kerr geodesic paths, and using the full energy spectrum of the continuum and reflection. From now on we will focus on the energy range of 0.3--10 keV, the band in which we will fit the data. Our binning is similar to the energy bins of the {\sc reflionx} model which is used to produce the time-averaged spectrum ($\approx 180$ bins from 0.3--10~keV). We measure the RRF in each energy band, $R(E)$, from the averaged spectrum, using the scale factor $R_{s}$ which is one of the model parameters. To investigate the general behaviours of the energy-dependent time lags, some of our model parameters are varied. The other parameters, when not stated, are kept constant at $h=5r_{g}$, $a=0.998$, $i=45^{\circ}$, $\Gamma=2$, $A=1$, $R_{s}=1$, $\xi_\text{ms}=10^4 \text{ erg cm s}^{-1}$, $p=2$ and the frequency range $0.9 - 3.6 \times 10^{-4} 1/t_{g}$.

\begin{figure*}
\centerline{
\includegraphics*[width=0.4\textwidth]{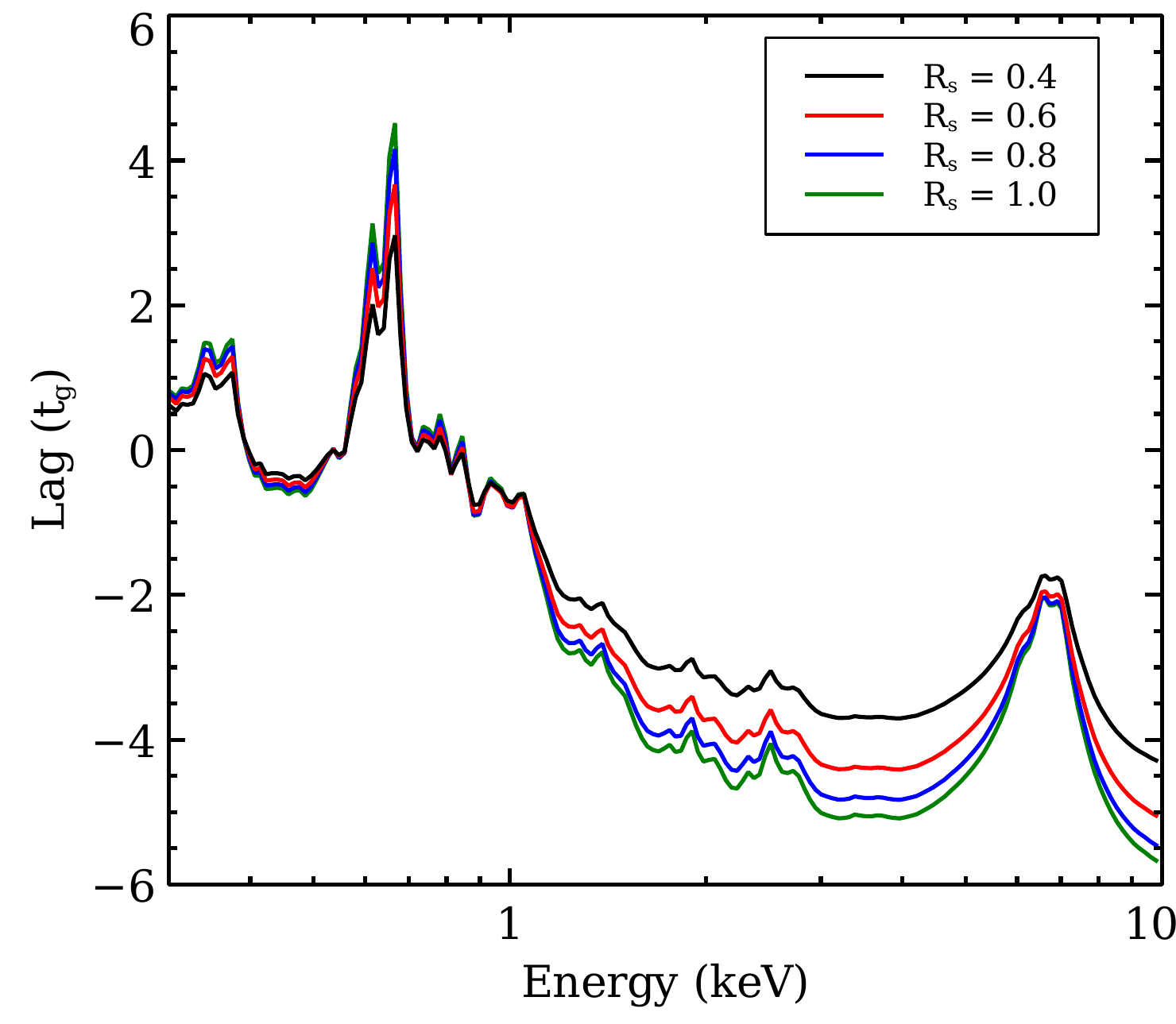}
\hspace{0.5cm}
\includegraphics*[width=0.4\textwidth]{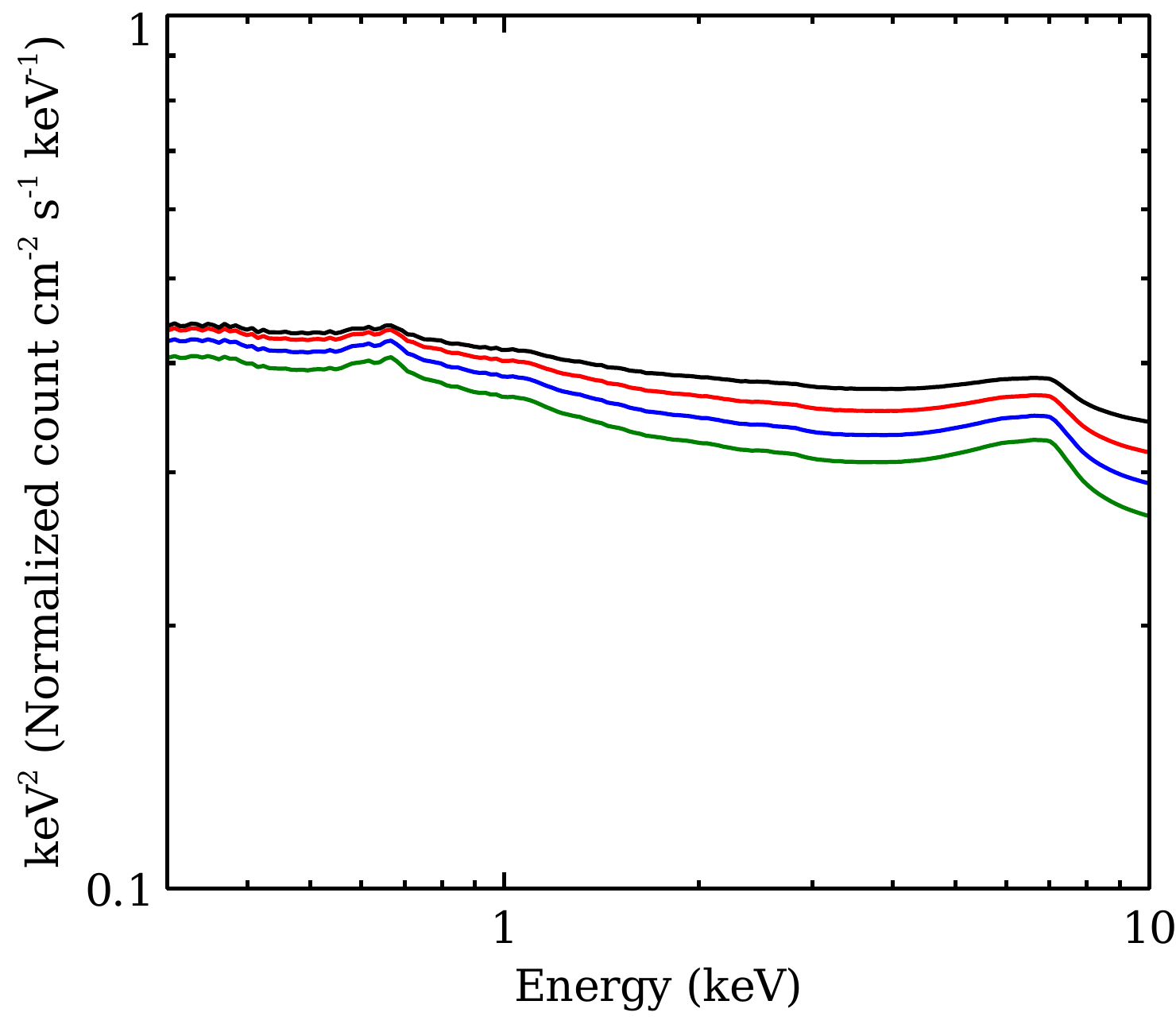}
}
\caption{Energy-dependent time lags (left) and time-averaged spectra (right) varying with $R_{s}$. The black, red, blue and green lines represent the cases when $R_{s}=$ 0.4, 0.6, 0.8 and 1, respectively. Increasing $R_{s}$ will decrease the power-law flux contaminated in the mean spectrum and hence time lags between the RDC and PLC increase.}
\label{lags-rs}
\end{figure*}

{\bf Scaling factor}. The scaling parameter, $R_{s}$, allows the RRF of all energy bands to change and hence affects the relative lags. Fig.~\ref{lags-rs} shows how the lag-energy and time-averaged spectra vary with $R_{s}$. Increasing $R_{s}$ will decrease the power-law flux contaminating the mean spectrum so stronger reflection features can appear. Since in this plot we fix the innermost region to be highly ionized, we are not seeing many spectral lines except the broad Fe K emission which is stronger for larger $R_{s}$. However, the effects of $R_{s}$ can be clearly seen in the lag-energy spectra as well as some of the spectral features from the X-ray reflection (e.g., the soft excess and the Fe K$\alpha$ line). As $R_{s}$ increases, so does the amplitude of the time delays of the Reflection Dominated Component (RDC; $\approx 0.3 - 1$ or $5- 7 \text{ keV}$ bands) behind the Power-Law Component (PLC; $\approx 1-4 \text{ keV}$ band). The reflection features that appear in lag-energy spectra for different values of $R_s$ are quite similar so the linear scaling-relation between the RRF and the lags \citep{Cackett2014,Uttley2014} should be applied, as long as the lags are probed at sufficiently low frequencies. The low-frequency maximum lag is set by the mean of the response function scaled by the RRF. However, when considering lag-energy profiles, each energy bin will have a unique value of RRF, $R(E)$, which is obtained by selecting $R_{s}$. This means that all energy bands have different factors in scaling their lags with respect to the reference band. Therefore, even though the changes within each band are linear with the model parameter $R_{s}$, the systematic changes across all bands, $R(E) \times R_s$, can be non-linear.

\begin{figure*}
\centerline{
\includegraphics*[width=0.4\textwidth]{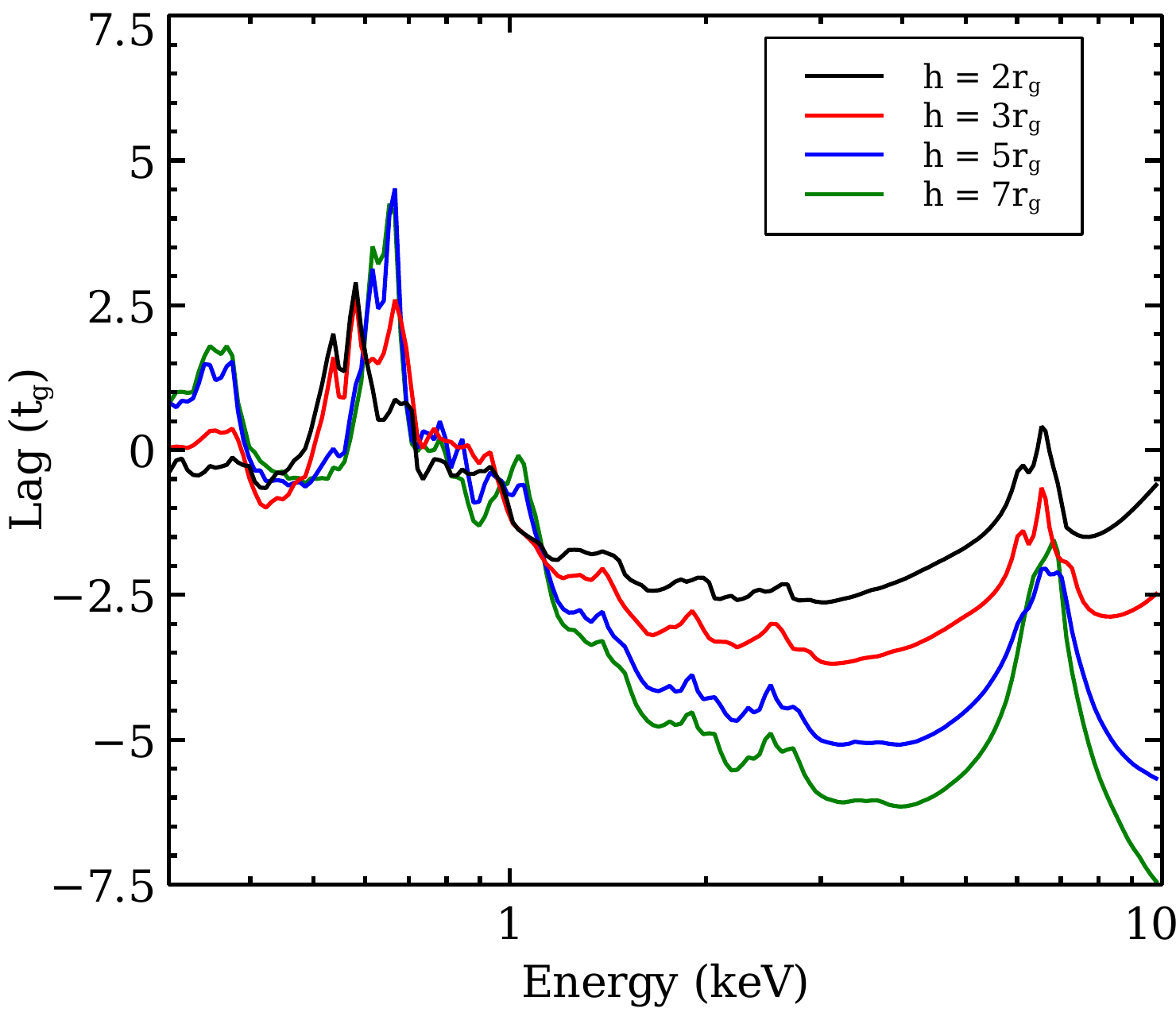}
\hspace{0.5cm}
\includegraphics*[width=0.4\textwidth]{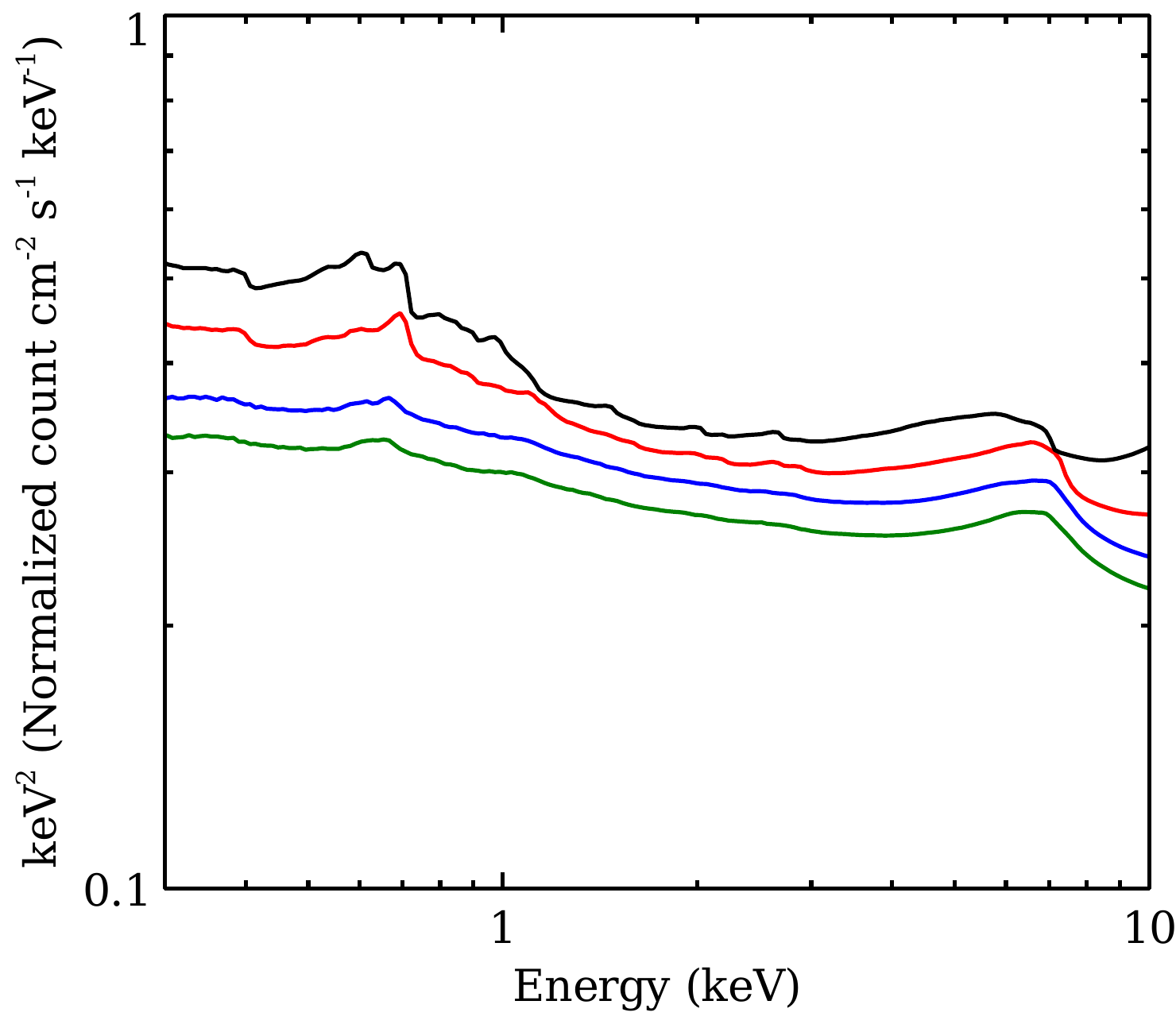}
}
\caption{Energy-dependent time lags (left) and time-averaged spectra (right) for the X-ray source located at $h=2r_g$ (black line), $3r_g$ (red line), $5r_g$ (blue line) and $7r_g$ (green line). The higher source height the longer light travel time of the reflection photons to the disc and the longer time lags between the PLC and RDC are measured.}
\label{lags-h}
\end{figure*}

{\bf Source height}. The time-averaged spectra along with the corresponding lag-energy spectra for different X-ray source heights are presented in Fig.~\ref{lags-h}. The disc is assumed to be highly ionized at the innermost part and less ionized further out depending on the illumination pattern and the density disc index which is fixed at $p=2$. We see in this case that greater heights lead to flatter time-averaged spectra since there are more photons illuminating large radii and then the disc can remain ionized across all outer annuli. The X-ray source closer to the black hole (e.g. $h=2r_g$) produces photons which are bent towards the centre meaning that the disc further away is less illuminated and becomes less ionized compared to the case of larger source height. This is why the characteristic features are more noticeable in the spectral profiles for lower source height. Even in this case the mean spectra are flatter for higher heights, with the opposite behaviour seen in the lag-energy profiles. Overall features are more enhanced with increasing source height because of longer mean response time. The larger source height produces longer time delays of the RDC behind the PLC which is consistent with the frequency-lag results \citep[see e.g.,][]{Wilkins2013,Cackett2014}. While changing $R_{s}$ preserves the features of the lag-energy spectrum, changing the source height affects the characteristic features of the profile, especially at soft energies and the $\approx 6.4$ Fe K$\alpha$ bands. Moreover, when the source height is larger than $\approx 5r_{g}$, the model predicts a clear sharp drop at $>7$ keV meaning that the 7--10 keV band significantly precedes the 2--4 keV continuum band. This signature has been observed in many AGN such as Mrk 335, Ark 546 and IRAS 13224-3809 \citep{Kara2013b, Kara2013a}. 

\begin{figure*}
\centerline{
\includegraphics*[width=0.4\textwidth]{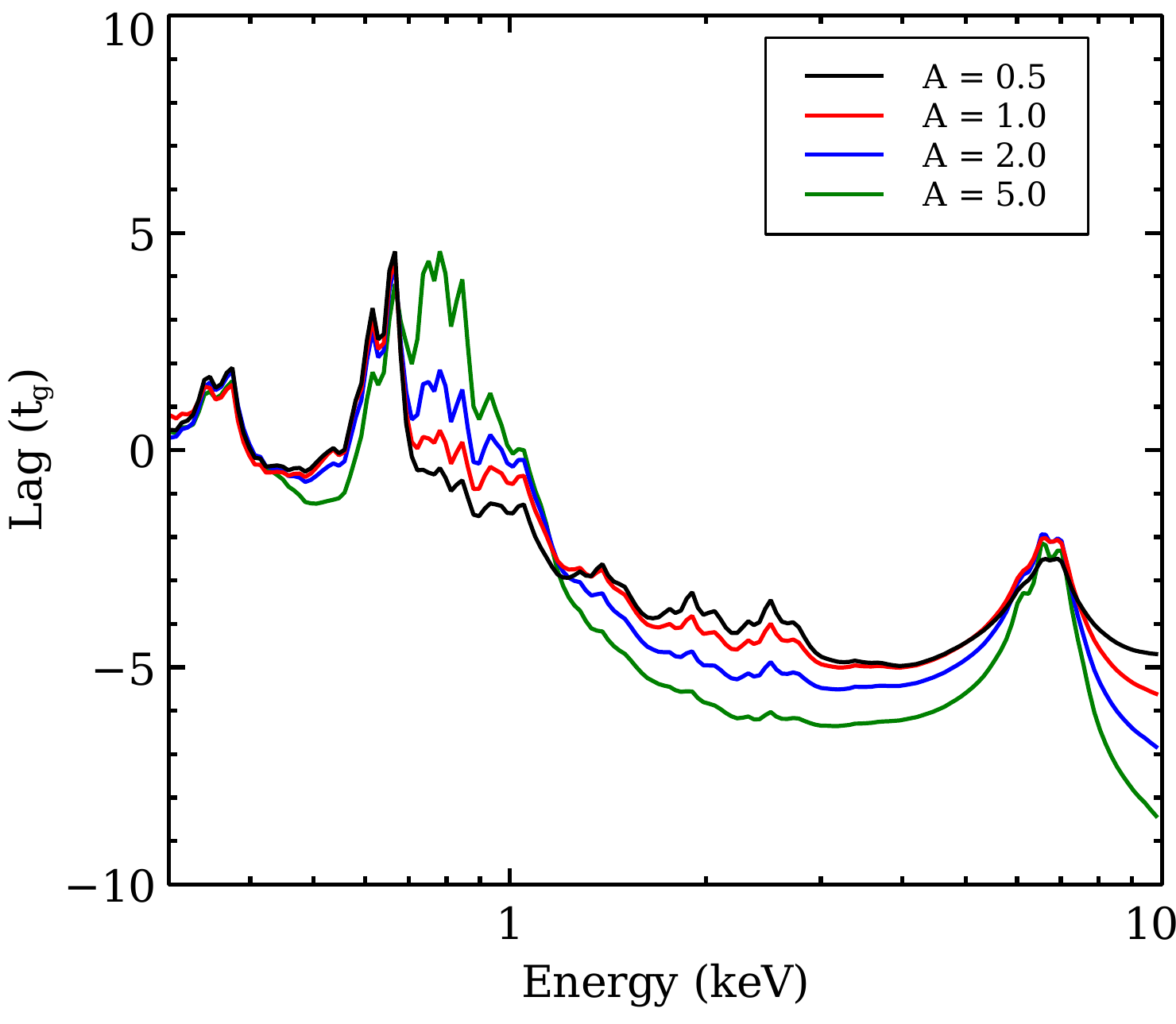}
\hspace{0.5cm}
\includegraphics*[width=0.4\textwidth]{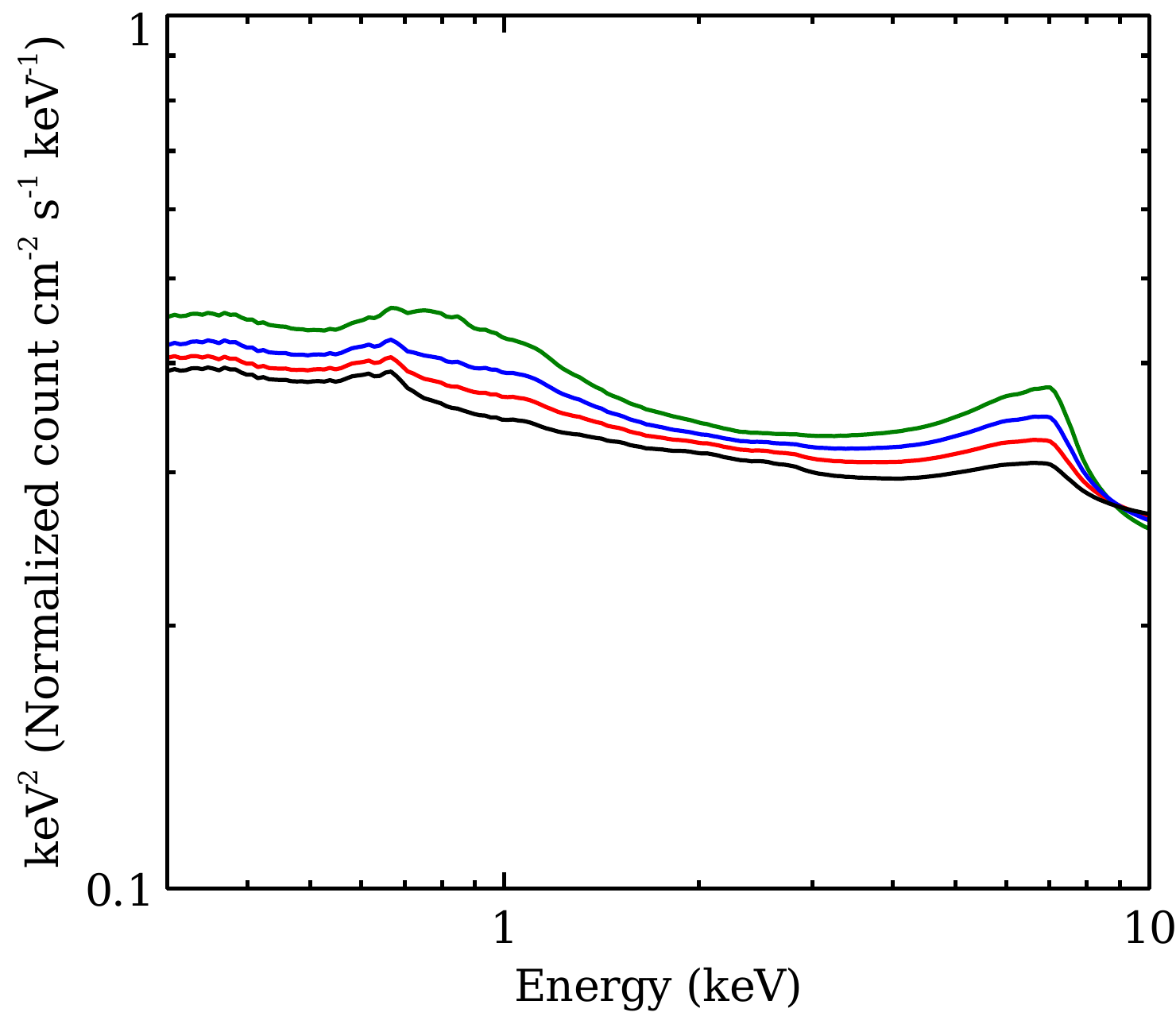}
}
\caption{Energy-dependent time lags (left) and time-averaged spectra (right) when the iron abundance, $A$, is varied. The black, red, blue and green lines represent the cases when $A=$ 0.5, 1, 2 and 5, respectively. The lag profile follows the trend of its spectral profile. For higher $A$, the lags between the RDC and PLC increase with lags of soft band gradually shift to higher energies.}
\label{lags-A}
\end{figure*}

{\bf Iron abundance}. Fig.~\ref{lags-A} shows that the lags between the RDC and PLC, and also the flux of the RDC bands in the spectral model, increase with the iron abundance, $A$. For high iron abundance we start to see the lags of soft band $\approx 0.6 \text{ keV}$ gradually shift to higher energies. A similar trend, but in shifting the soft-band flux rather than the lags, has been found in the rest-frame emission spectrum with variable iron abundance \citep[see e.g.,][]{Ross2005}. We show that these features after blurring are also noticeable in both the time-averaged spectrum and the corresponding lag profile. It is therefore obvious that the energy dependent reverberation lags, theoretically, follow the changes of the spectral flux as the RRF measured from the spectrum plays a role in determining the lags. It should be noted that the high iron abundance also enhances the contrast between the lag in the iron line and the 7--10~keV tail.

\begin{figure*}
\centerline{
\includegraphics*[width=0.4\textwidth]{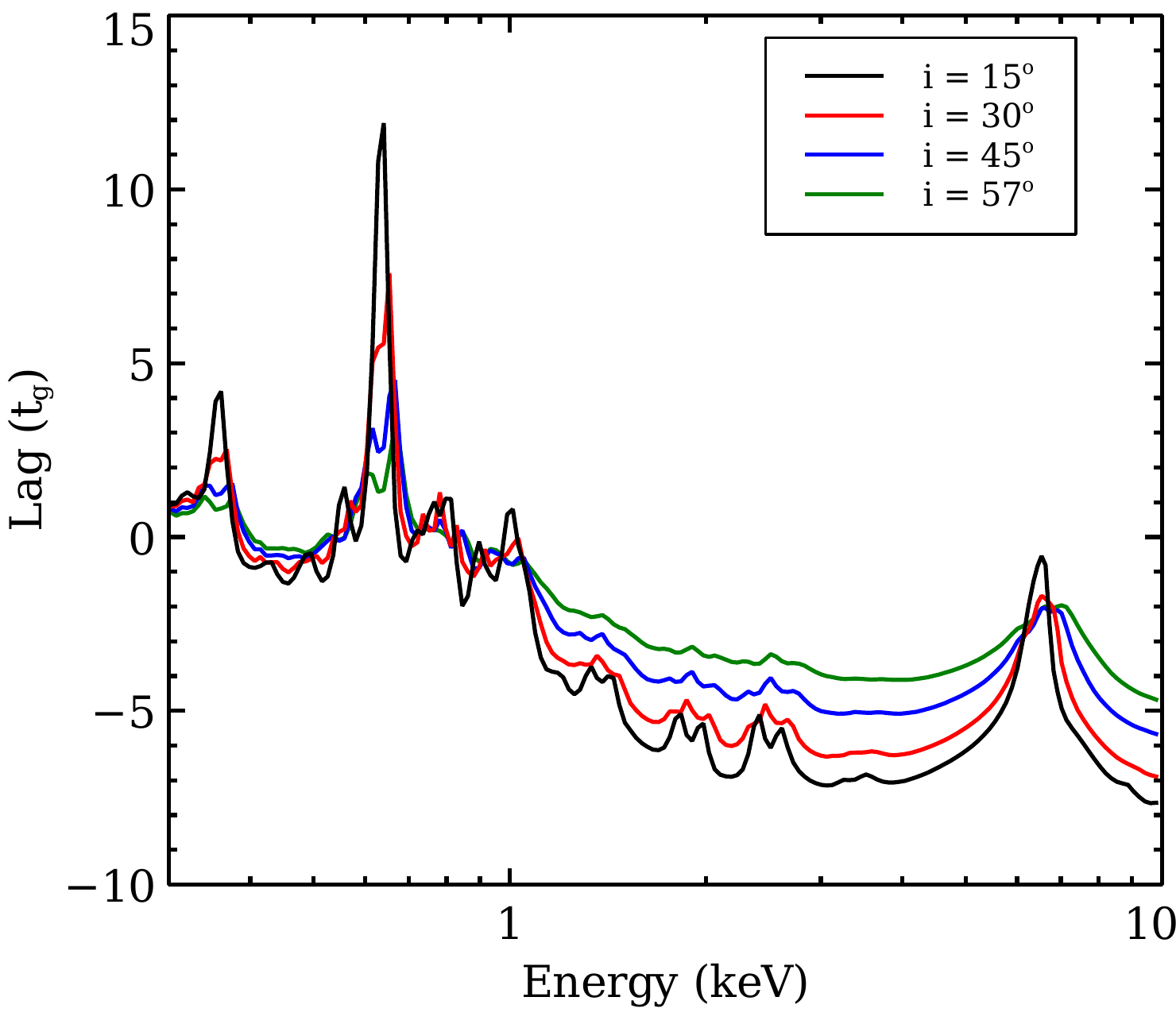}
\hspace{0.5cm}
\includegraphics*[width=0.4\textwidth]{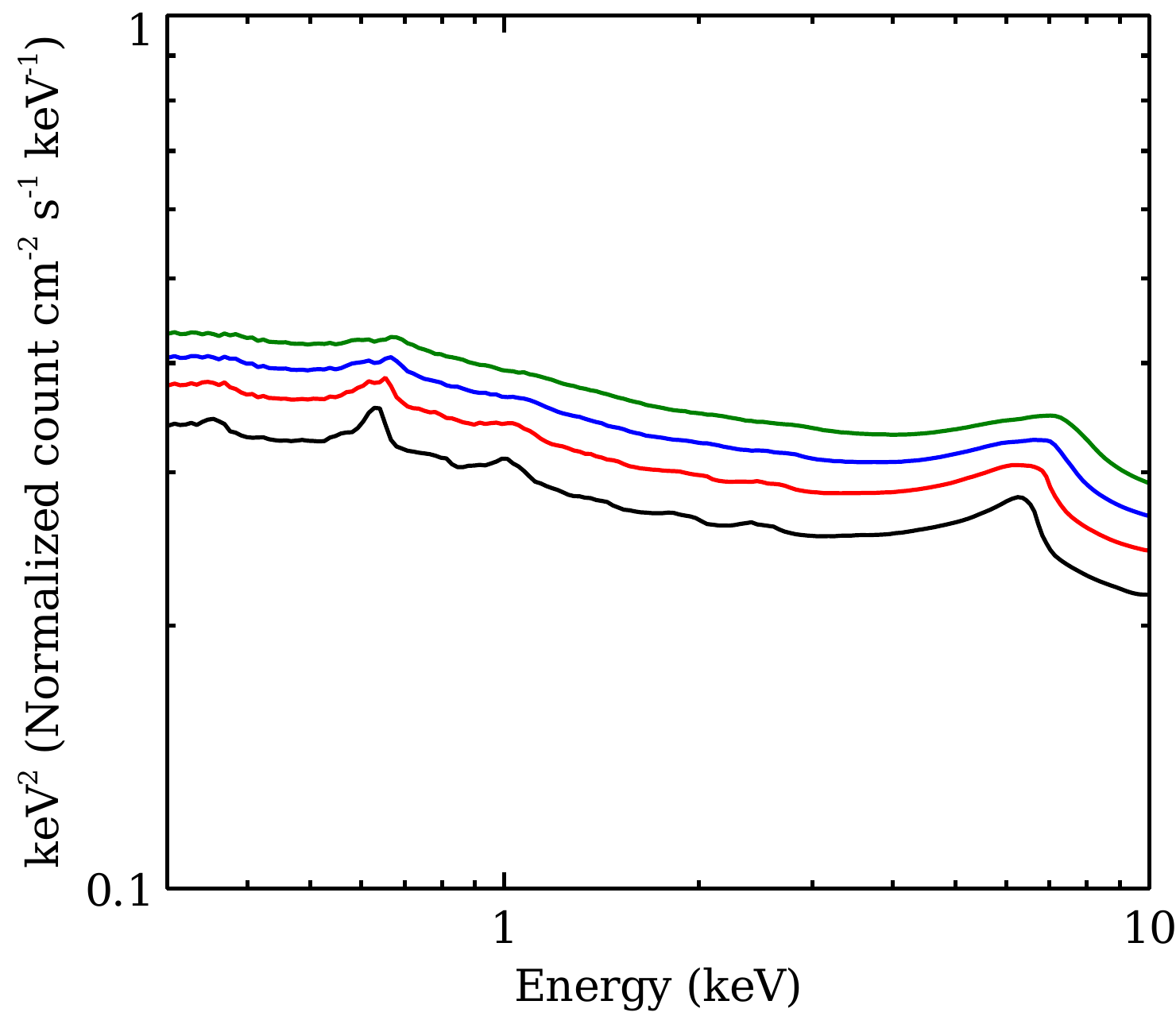}
}
\caption{Energy-dependent time lags (left) and time-averaged spectra (right) varying with an inclination $i$. The black, red, blue and green lines represent the cases when $i=15^{\circ}$, $30^{\circ}$, $45^{\circ}$ and $57^{\circ}$, respectively. The more edge-on the disc is viewed, the more the Fe K lags are broadened but the less time delays between the RDC and PLC are measured.}
\label{lags-i}
\end{figure*}

{\bf Inclination angle}. How the lags varies with the inclination angle, $i$, follows the trend of how the inclination affects the mean spectrum (Fig.~\ref{lags-i}). The more edge-on the disc is viewed, the more the Fe K lags are broadened and the blue wing of the lags are shifted to the higher energies. For higher inclination, the reflection photons from the near (far) side of the disc will take shorter (longer) time to the observer. Increasing inclination then increases a chance of detecting reflection photons at a broader range of time (i.e., broader response function). We find the lags between the RDC and the PLC decrease with increasing inclination, in agreement with the results found by modelling the lag-frequency spectrum \citep{Cackett2014,Emmanoulopoulos2014}.

\begin{figure*}
\centerline{
\includegraphics*[width=0.4\textwidth]{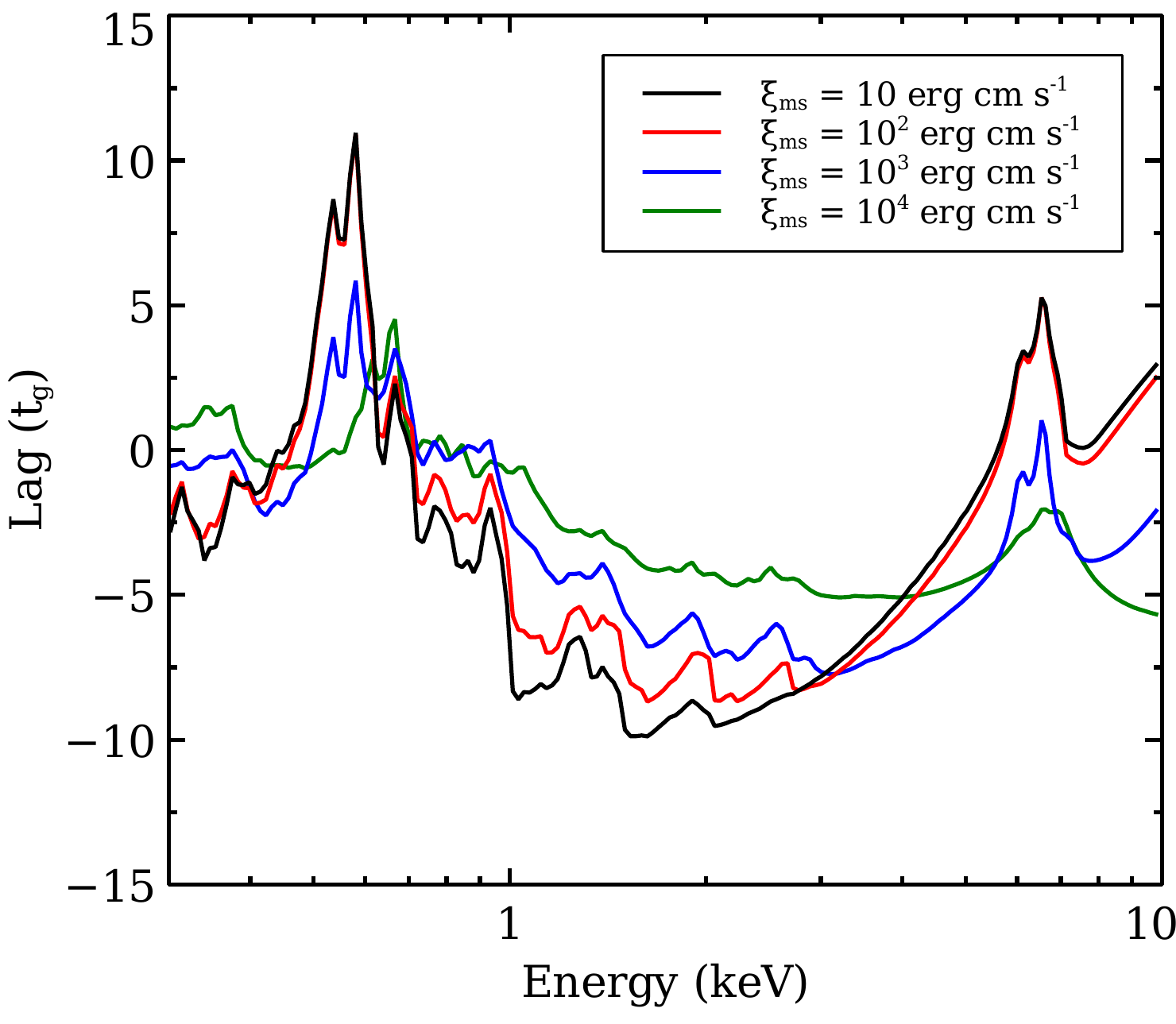}
\hspace{0.5cm}
\includegraphics*[width=0.4\textwidth]{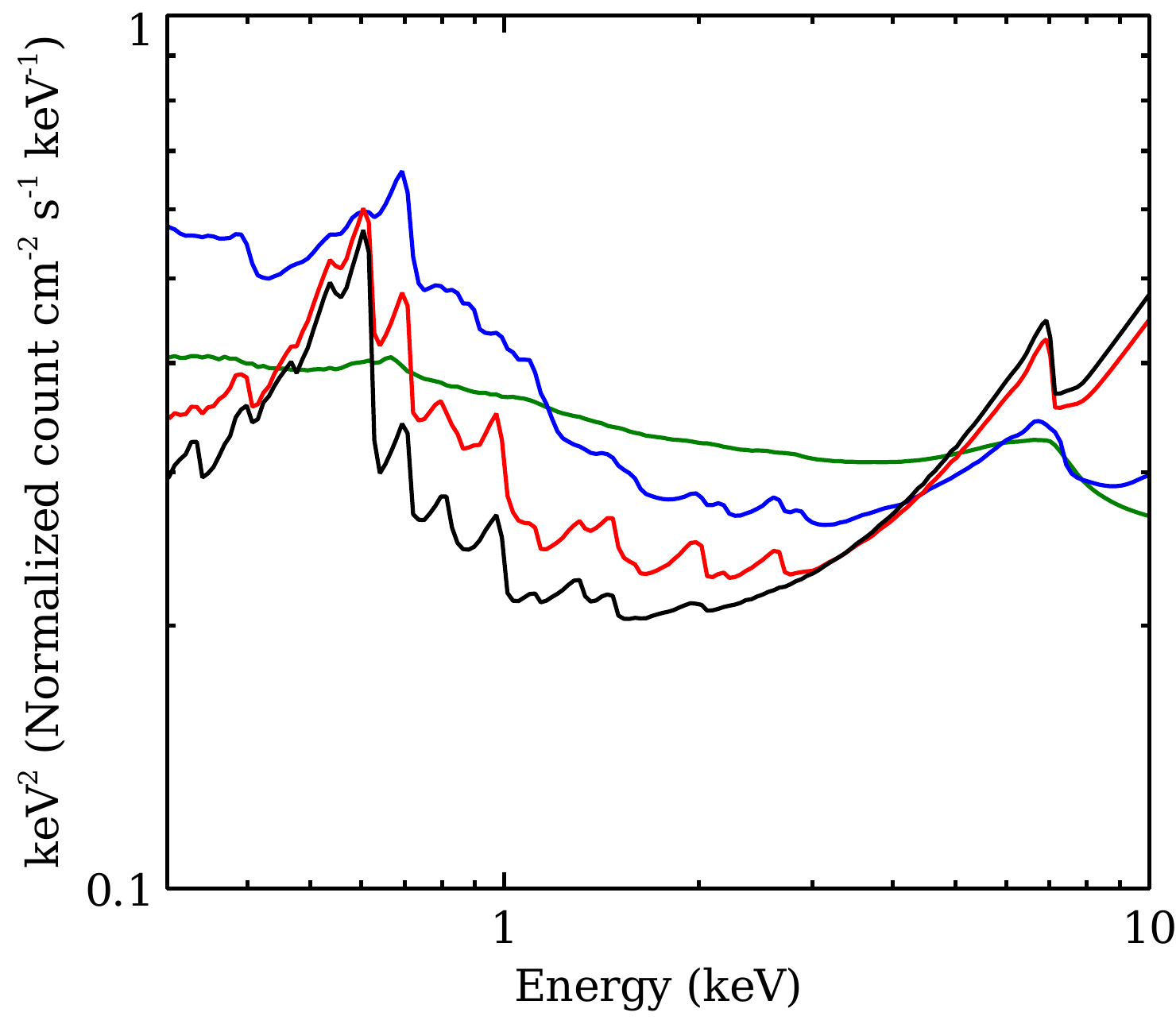}
}
\caption{Energy-dependent time lags (left) and time-averaged spectra (right) for the ionization parameter at the innermost region $\xi_\text{ms}=10 \text{ erg cm s}^{-1}$, (black line), $10^{2} \text{ erg cm s}^{-1}$ (red line), $10^{3} \text{ erg cm s}^{-1}$ (blue line) and $10^{4} \text{ erg cm s}^{-1}$ (green line). See text for more details.}
\label{lags-xi}
\end{figure*}

{\bf Ionization parameter}. A high ionization parameter at the inner radius gives a flatter reflection spectrum which decreases the relative lags between the RDC and the PLC because they are more diluted by the cross-components (Fig.~\ref{lags-xi}). This also agrees with what is reported in \cite{Chainakun2015} by looking at the effects of the ionization gradient in the disc on the frequency-dependent time lags. Note that our value of the ionization parameter, $\xi_\text{ms}$, is not the value for the entire disc. We define the $\xi_\text{ms}$ at the $r_\text{ms}$ ($\approx 1.235 r_g$ for $a=0.998$) and the ionization parameter further out is determined by the illumination pattern and the disc density index, $p$. As predicted by our model, the leading of $\approx$ 7--10~keV band ahead of the continuum dominated band also requires the disc to be highly ionized at the innermost part, in addition to the high source height which is required as mentioned before. 

Our model predicts the 3~keV dip in the lag-energy spectrum as seen in many AGN \citep[e.g.,][]{Kara2013b,Kara2013a} under the same condition as the drop of the lags in the 7--10~keV band. Time leads or lags depend on the response of each band and how much the reflection flux contributes to that band on top of the continuum flux. The reflection photons detected at $\approx 3 \text{ keV}$ are likely from the red-shifted innermost region, but if the inner disc is highly ionized this region would give very small reverberation lags because of the flat emergent spectrum. The lack of time-lags from the inner, ionised disc allows the possibility of the 3~keV band leading the reference band as the spectra from outer annuli, being less ionized, will have lower flux around 3~keV. These spectra then cause the 3~keV band to lead the adjacent energy bands. The different responses in each energy bin are not only due to the redshifts experienced by the reflection from the different regions but also due to changes in the reflection itself due to the ionization gradients in the disc. This means that it makes a significant difference if we treat the disc as having a single ionization parameter. So far the traditional 3~keV dips and the 7--10 keV band preceding the continuum can be produced only in limited cases, in which the source height is $> 5 r_g$ and the disc is highly-ionized near the centre and less ionized further away. Nevertheless, the outer radii should not be totally cold ($h_{s}>5r_{g}$ is ensuring that the outer part of the disc is still sufficiently irradiated), otherwise the reflection flux and the RRF at $>7$~keV will increase significantly (see Fig.~\ref{lags-xi}) leading to smaller lag-differences between this band and Fe K band.

\begin{figure*}
\centerline{
\includegraphics*[width=0.4\textwidth]{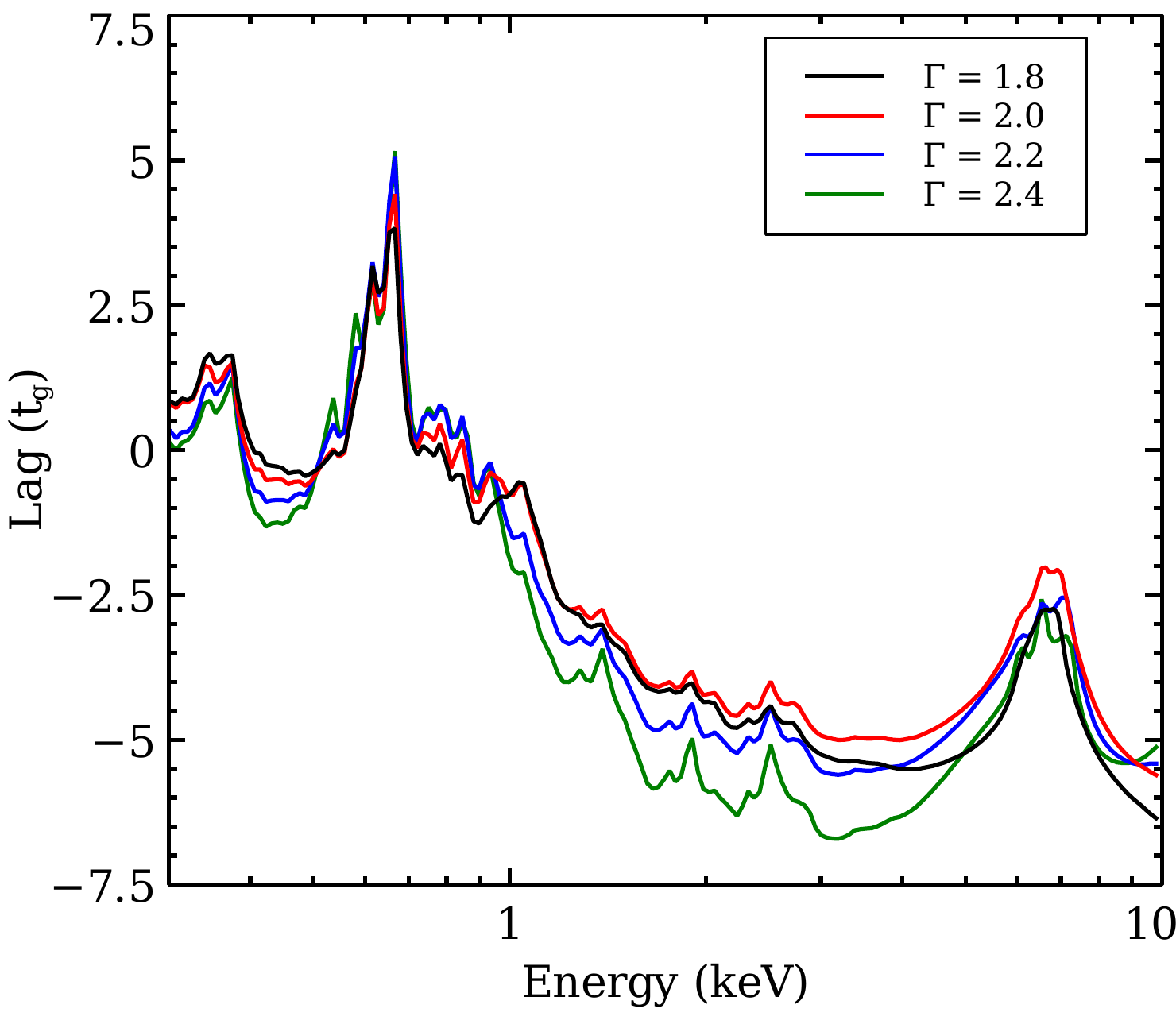}
\hspace{0.5cm}
\includegraphics*[width=0.4\textwidth]{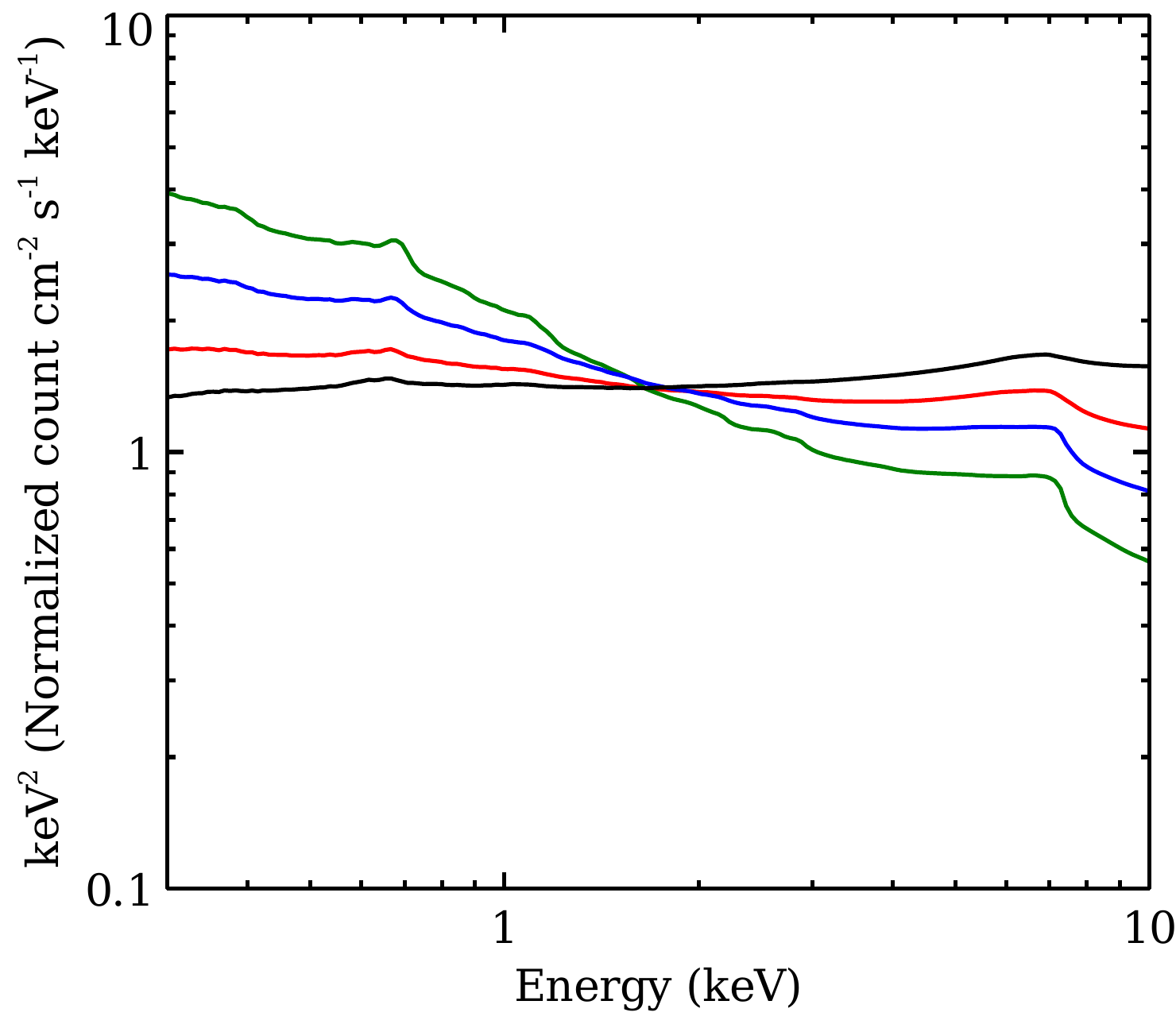}
}
\caption{Energy-dependent time lags (left) and time-averaged spectra (right) when the photon index $\Gamma=1.8$ (black line), 2.0 (red line), 2.2 (blue line) and 2.4 (green line). See text for more details.}
\label{lags-g}
\end{figure*}

{\bf Photon index}. From Fig.~\ref{lags-g}, the photon index, $\Gamma$, has a significant effect on the time-averaged spectrum but has much less effect on the lags. This is because the slope of the mean spectrum changes with the photon index, its constituent parts which are the continuum and reflection spectrum also have their slopes changed accordingly hence the mean RRF is almost identical for all $\Gamma$. Therefore the spectral model, rather than the timing model, is preferable to constrain the photon index from the observational data. 

\begin{figure}
\centering  
\includegraphics*[width=70mm]{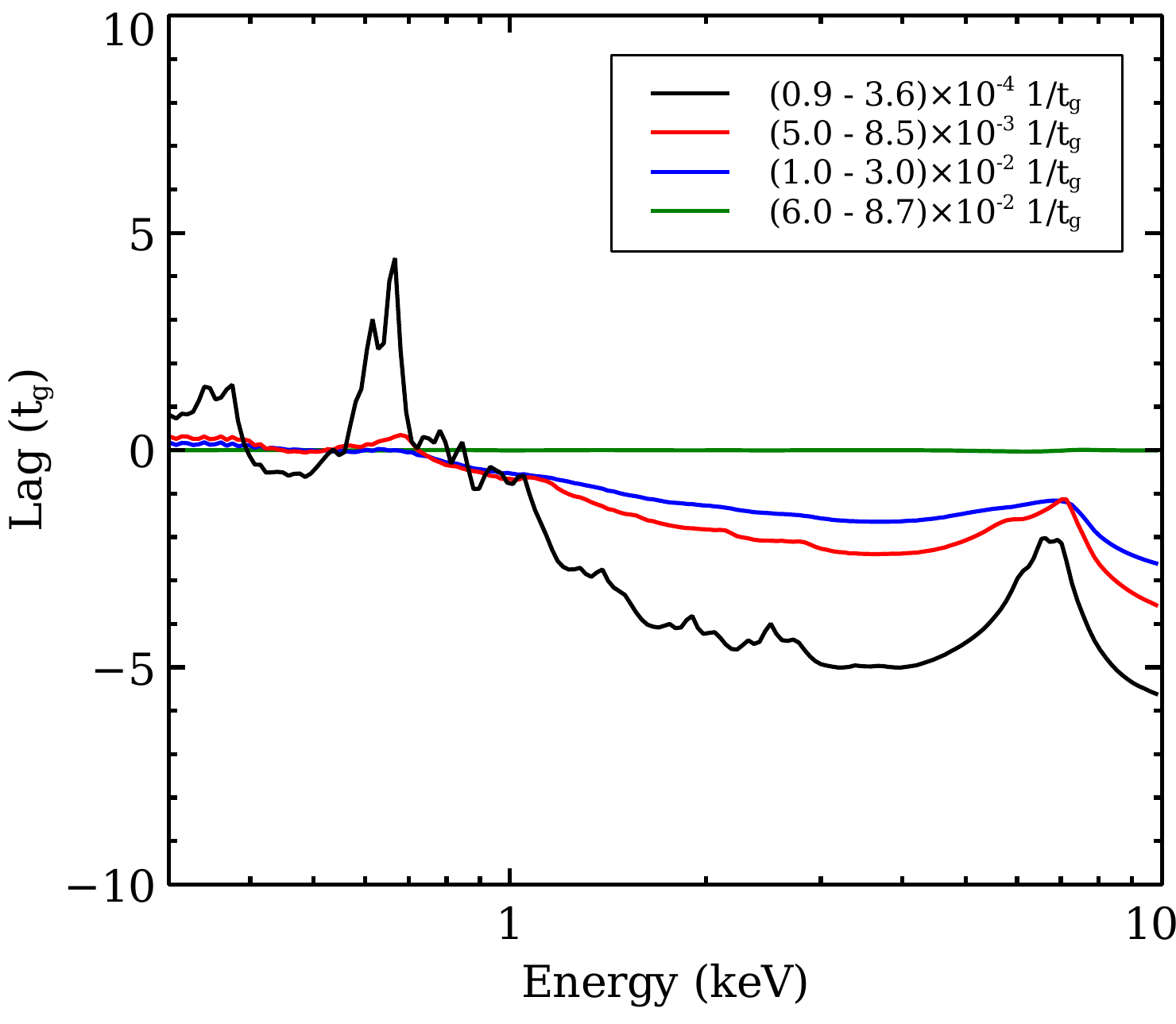}
\caption{Energy-dependent time lags when the selected frequency ranges are 0.9--3.6$\times 10^{-4} 1/t_{g}$ (black line), 5--8.5$\times 10^{-3} 1/t_{g}$ (red line), 1--3$\times 10^{-2} 1/t_{g}$ (blue line) and 6--8.7$\times 10^{-2} 1/t_{g}$ (green line). \label{lags-f}}
\end{figure}

{\bf Frequency range}. The specific frequency range over which we are modelling the reverberation lags is also important. Higher frequencies mean that shorter time-scales on the disc are probed. At the lowest frequencies (i.e., longest time-scales) the reverberation signature is probed across the entire disc. For higher frequencies, only the inner and more red-shifted region is probed so we see the lag-energy profiles are relatively broader as the outer part and more blue-shifts part are cut-out (Fig.~\ref{lags-f}). The lag-energy spectrum, similarly to the lag-frequency spectrum, should be analyzed only at frequencies less than the phase-wrapping frequency, $f_{w}$. The interpretation of the lags at frequencies $>f_{w}$, as we know so far, is meaningless. 

\begin{figure}
\centering  
\includegraphics*[width=70mm]{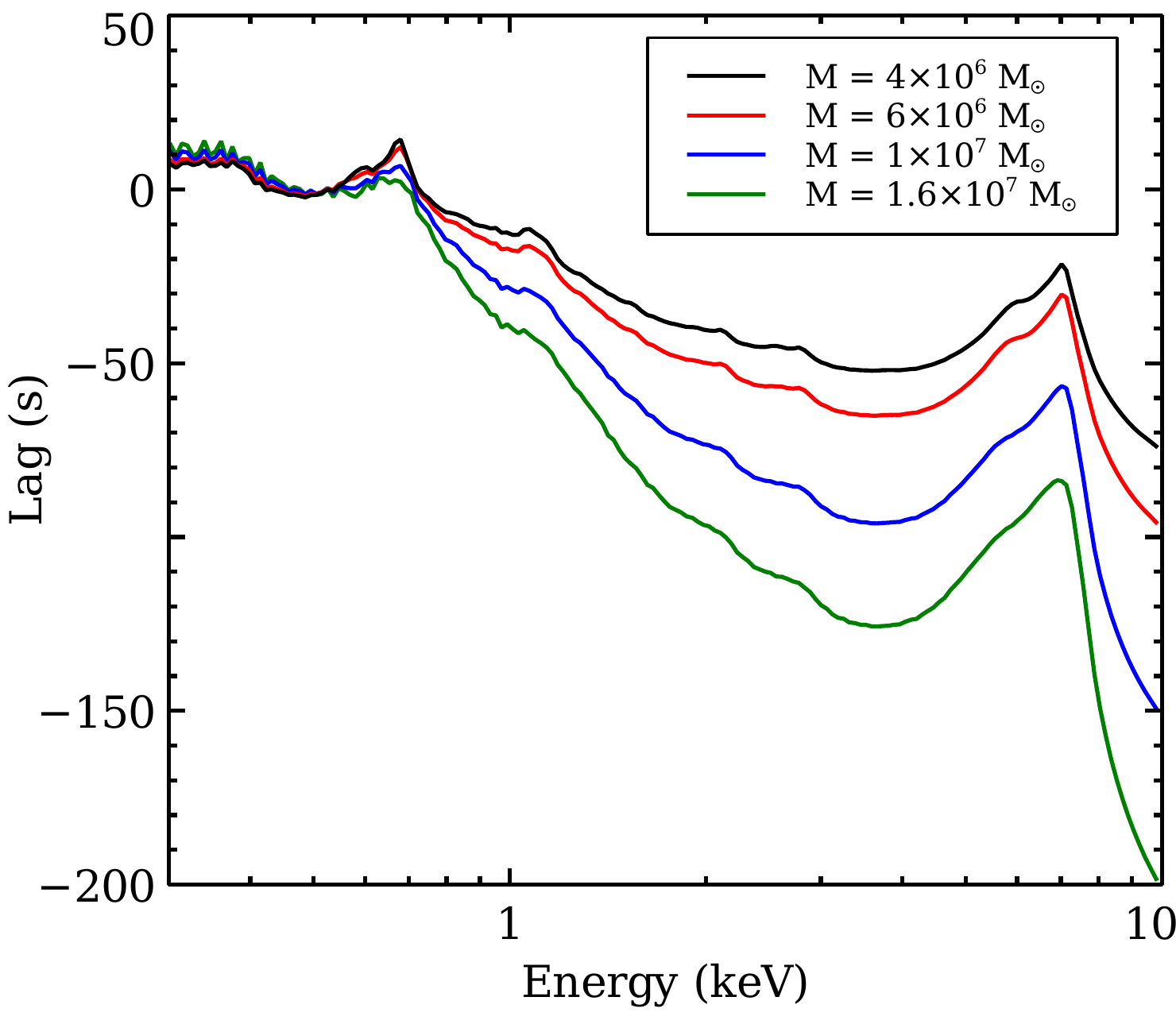}
\caption{Energy-dependent time lags scaling with the black hole mass of $4 \times 10^{6} M_{\odot}$ (black line), $6 \times 10^{6} M_{\odot}$ (red line), $ 10^{7} M_{\odot}$ (blue line) and $1.6 \times 10^{7} M_{\odot}$ (green line). \label{lags-M}}
\end{figure}

{\bf Black hole mass}. Assuming a black hole mass, $M$, the lags in the geometrical units can be transformed to physical units (e.g., for $M=10^{6}M_{\odot}$, $1t_{g} \approx 5$s). Fig.~\ref{lags-M} shows how the lag-energy spectrum scales with $M$. The different lags between the RDC and the PLC increase with $M$, but it should be noted that as we fix the frequency range of interest the phase wrapping will occur at lower frequencies for higher values of $M$, as the frequencies are scaled with a factor $1/t_{g}$. Therefore there is a limit to $M$ that sets the largest time-lag differences in the lag-energy spectrum for each frequency range. Beyond that limit the situation is reversed in the sense that increasing $M$ will suppress the relative lags in the lag-energy profile.

For the final investigation, we change the reference band and see how this affects the lags (the result is not presented here). We find that, the model provides the similar relative lags independent on the choices of reference bands. This is what we expect and has been guided by the observation.

\section{Fitting procedure}

We adapt the technique presented by \cite{Chainakun2015} where the time-averaged spectrum and the frequency-dependent time lags are simultaneously fit using {\sc isis} \citep{Houck2000}. We define two separate data sets which are the observed time-averaged and lag-energy spectra. Both data sets have been transformed so they are stored internally as arrays of binned counts vs. wavelength. We first produce a course global grid of model parameters that covers all possible values reported in previous studies of the three AGN, Mrk 335, Ark 564 and IRAS 13224-3809. For each grid cell the modelled time-averaged spectrum and energy-dependent time lags are fitted to the corresponding data-set, with the parameters of additional model components minimised using the {\sc subplex} algorithm to estimate the $\chi^2$ statistic at each point in the grid. We step through each grid cell and note the best fitting model, i.e., the one with the lowest $\chi^2$ value. We then expand the model by producing finer, local grids around that particular course grid cell. The fitting procedure is then repeated with those local grids to find the best, and more precise, fitting parameters.

It is worth mentioning that when fitting the data in {\sc isis} the background is always added to the model instead of being subtracted from the data. The error bars are modified to account for the statistical uncertainty on the background. This is not optional but the resulting $\chi^2$ value is confirmed to be identical if the same data and model are fit using {\sc xspec} \citep{Arnaud1996}.

\section{Results}

The spectral- and timing-model expression for our fits are presented in Table~\ref{tab:model_expression}. For Mrk 335, the blurred component of the time-averaged spectrum is produced by our model ({\sc revb}). We include the unblurred-{\sc reflionx} and {\sc mekal} models to account for distant reflection from neutral gas and thermal emission from an ionized plasma, respectively. These components were identified in the \emph{XMM-Newton} spectrum by \citet{Oneill2007}. The warm absorber and the disk thermal emission are modelled by {\sc xstar} \citep{Kallman2001} and {\sc blackbody}, respectively. The total spectrum is modified by the galactic absorption ({\sc tbabs}) set to be 3.7, 5.3 and 6.5 $\times 10^{20} \text{ cm}^{-2}$ for Mrk 335, IRAS 13224-3809 and Ark 546, respectively. For the timing model, besides the reverberation lags ({\sc revb}), we include the power-law lags ({\sc powerlaw}) that increase with increasing energy. These lags are plausibly produced by the propagation of accretion rate fluctuation inwards through the disc \citep[e.g.,][]{Arevalo2006} and are expected even at high frequencies where the reverberation lags dominate. We do not know in advance whether or not the power-law lags also contribute so both slope and normalization of this component are allowed to be free parameters. In cases of IRAS 13224-3809, the model components are similar to those of Mrk 335 except that the thermal component from the ionzied gas is neglected. The spectrum of Ark 564, however, requires an additional broad Gaussian line at the soft excess band with the centroid energy $\approx 0.4$~keV. This component is broad so it should contribute to both spectral and timing models as its origin should be the inner-disc reflection. More specifically, we find that the geometry of Ark 564 is complex in the way that the warm absorber covers only the distant reflection component.  

\begin{table*}
\begin{tabular}{ l l l }
\hline
\multirow{2}{*}{AGN} & \multirow{2}{*}{Spectral-model expression} &  \multirow{2}{*}{Timing-model expression} \\  
 \\ \hline

Mrk 335 & $\textsc{tbabs} \otimes \textsc{xstar1} \otimes  (\textsc{revb} + \textsc{reflionx} + \textsc{mekal} + \textsc{diskbb})$ & $\textsc{revb} + \textsc{powerlaw}$  \\
IRAS 13224-3809 & $\textsc{tbabs} \otimes \textsc{xstar1} \otimes  \textsc{xstar2}\otimes  (\textsc{revb} + \textsc{reflionx} +  \textsc{diskbb})$ & $\textsc{revb} + \textsc{powerlaw}$  \\
Ark 564 & $\textsc{tbabs} \otimes   (\textsc{revb} + \textsc{xstar1} \otimes  \textsc{xstar2}\otimes (\textsc{reflionx}) +  \textsc{diskbb} + \textsc{zgauss}$) & $\textsc{revb} + \textsc{powerlaw}  + \textsc{zgauss}$  \\

\hline
\end{tabular}
\caption{The spectral- and timing-model expression for the fits of three AGN. It should be noted that adequate fits can still be achieved without the power-law lag component, {\sc powerlaw}.	 \label{tab:model_expression}}
\end{table*}

\begin{figure*}
\centerline{
\hspace{0.2cm}
\includegraphics*[width=0.55\textwidth]{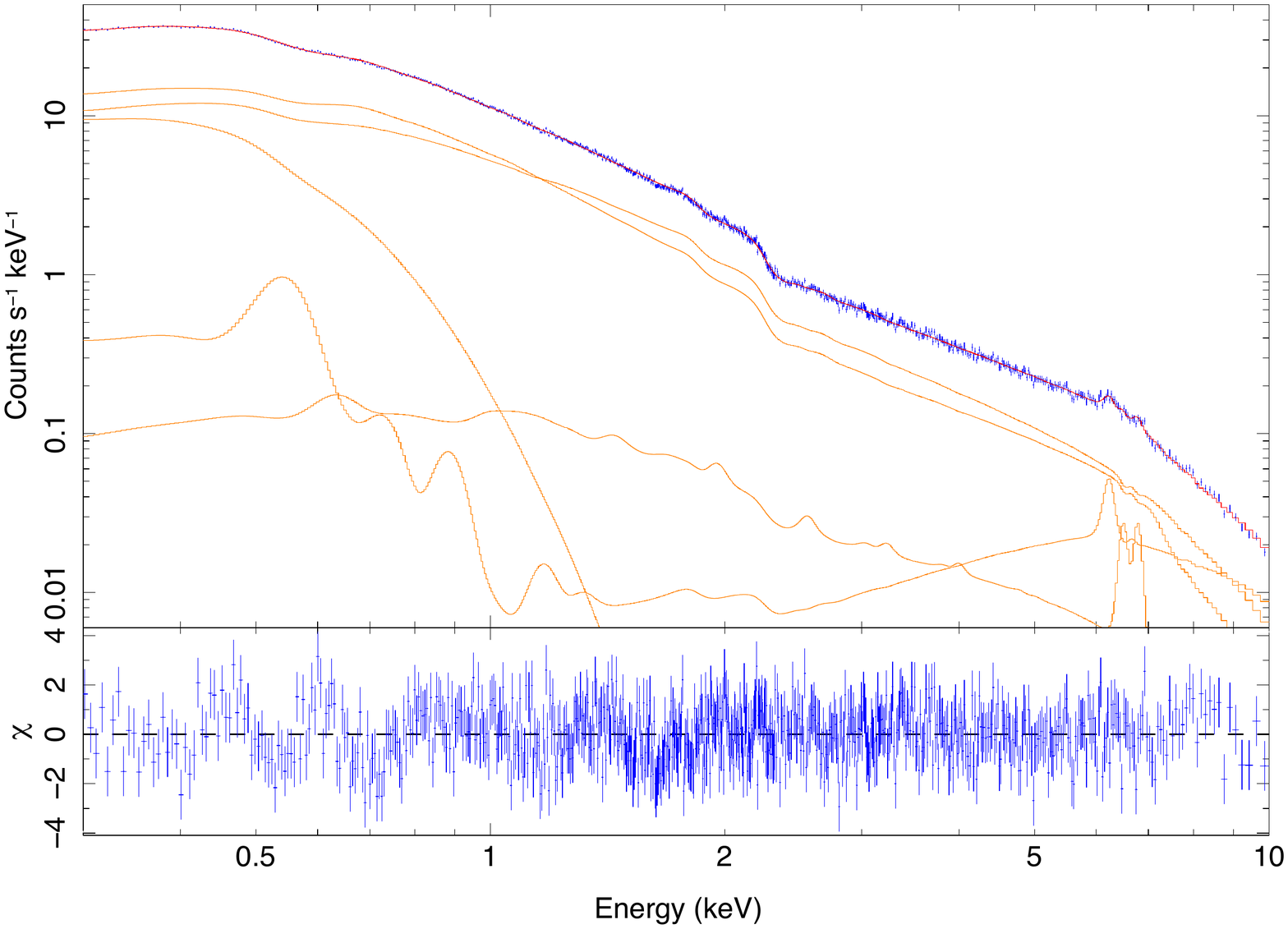}
\hspace{-0.5cm}
\includegraphics*[width=0.50\textwidth]{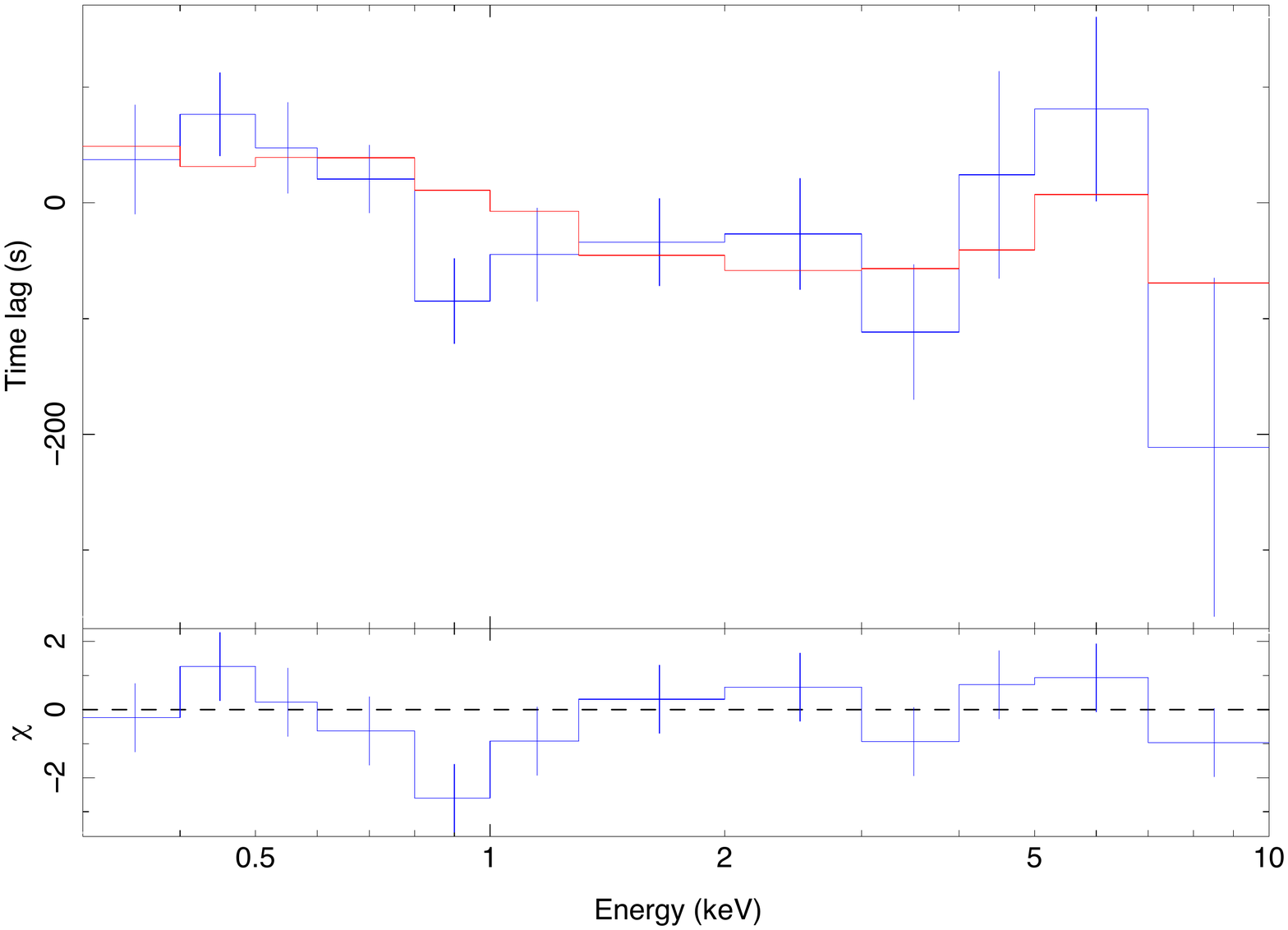}
\vspace{-1.0cm}
}
\centerline{
\hspace{0.2cm}
\includegraphics*[width=0.55\textwidth]{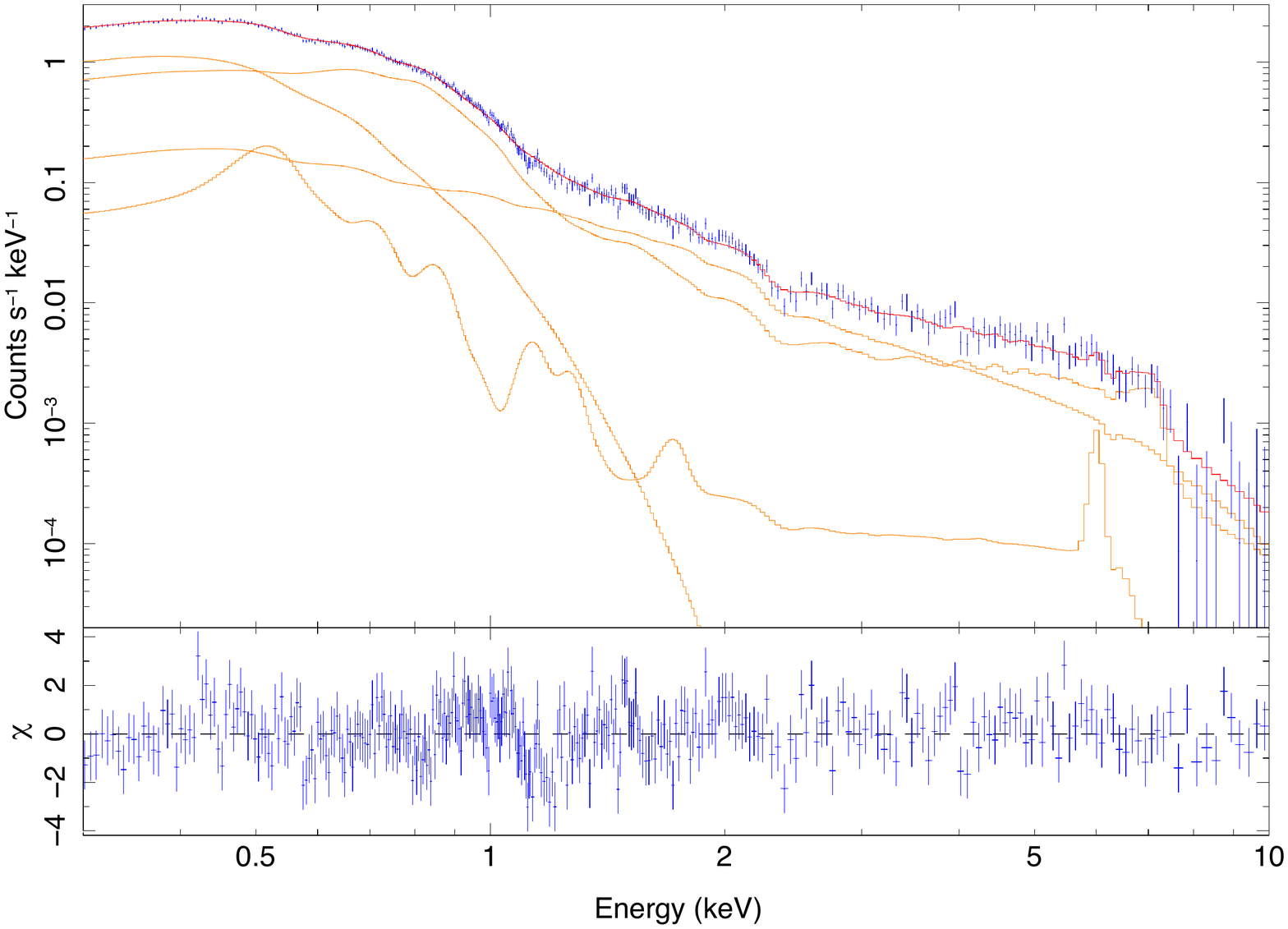}
\hspace{-0.5cm}
\includegraphics*[width=0.50\textwidth]{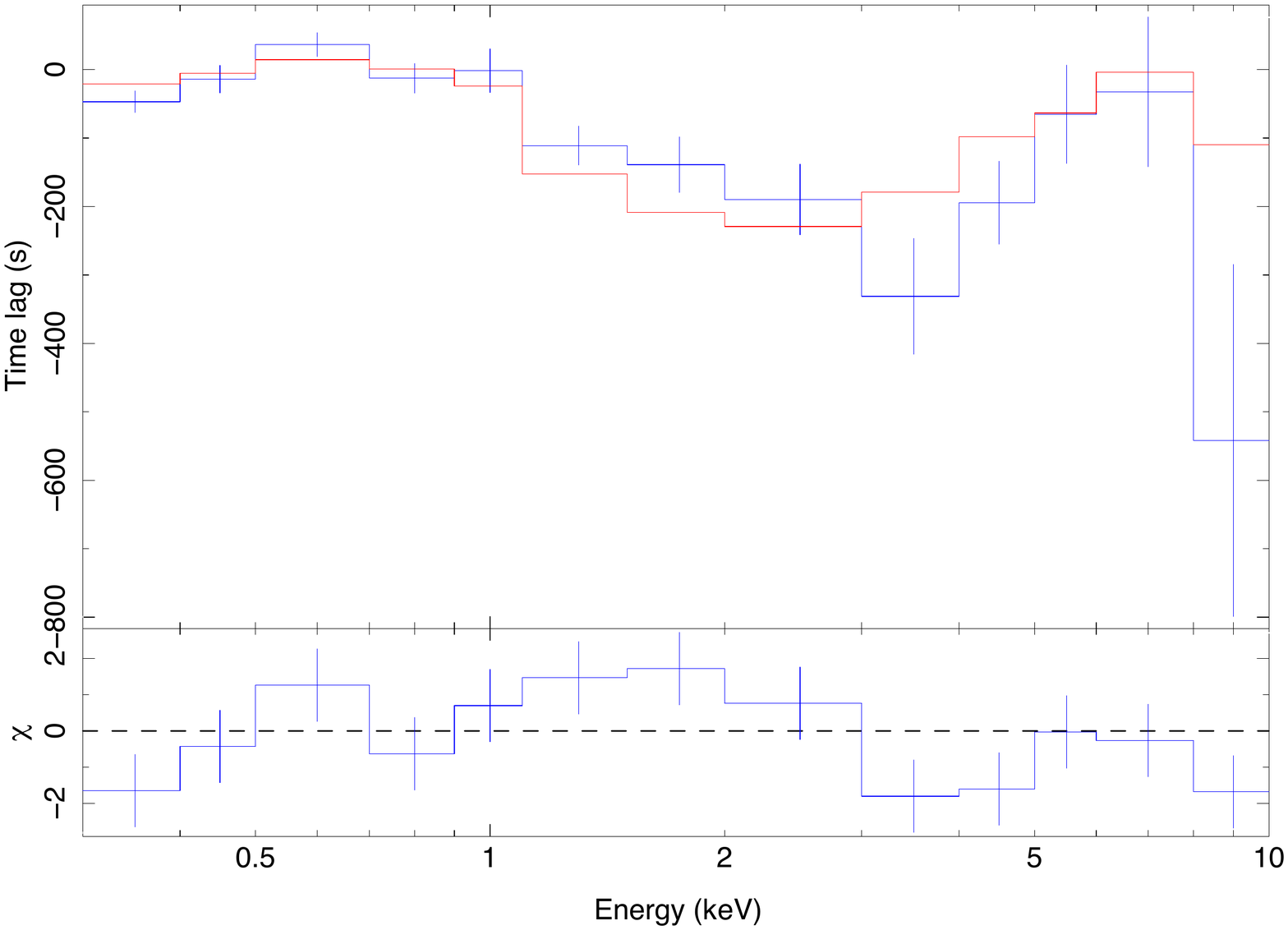}
\vspace{-1.0cm}
}
\centerline{
\hspace{0.56cm}
\includegraphics*[width=0.55\textwidth]{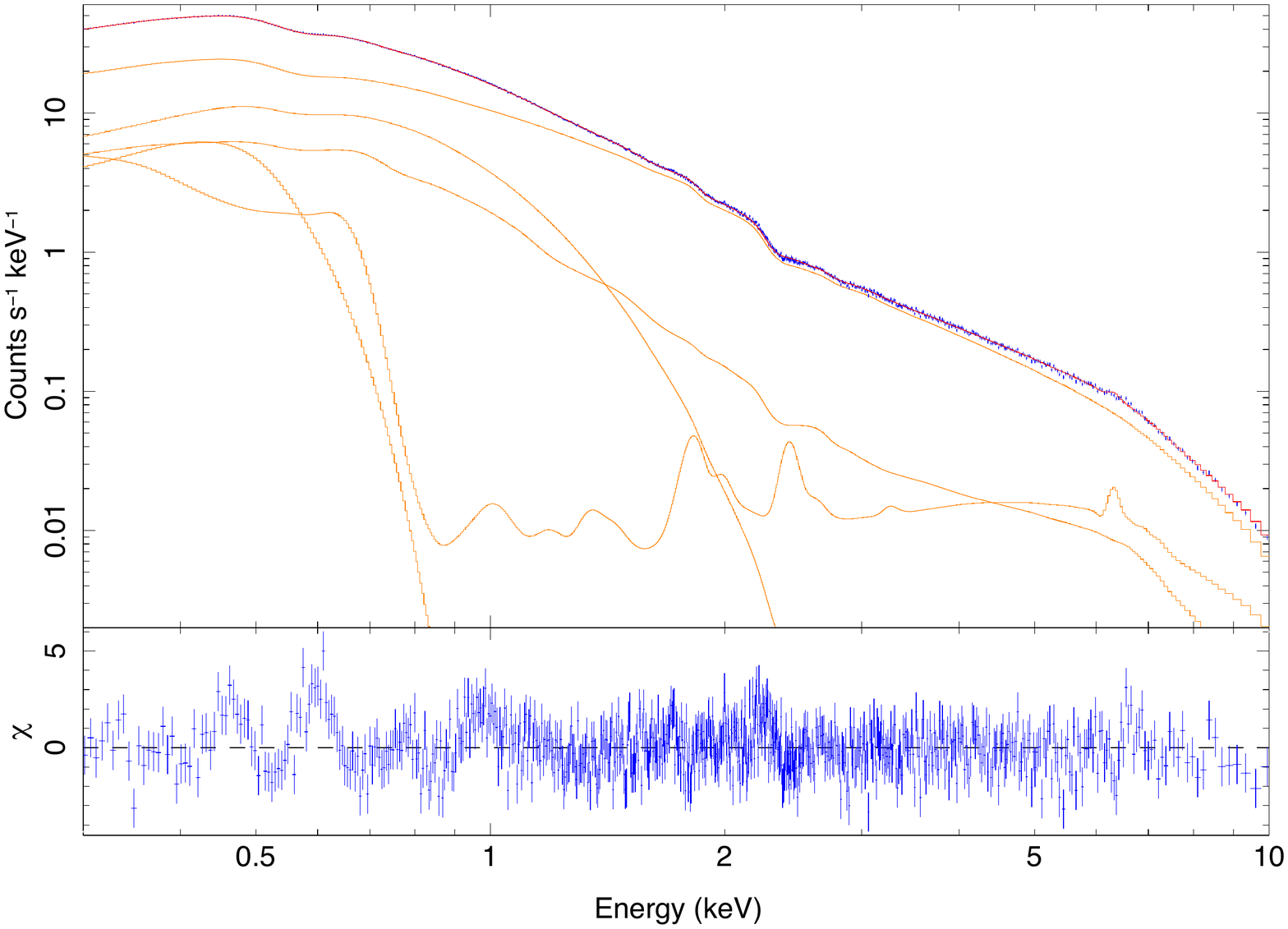}
\hspace{-0.97cm}
\includegraphics*[width=0.55\textwidth]{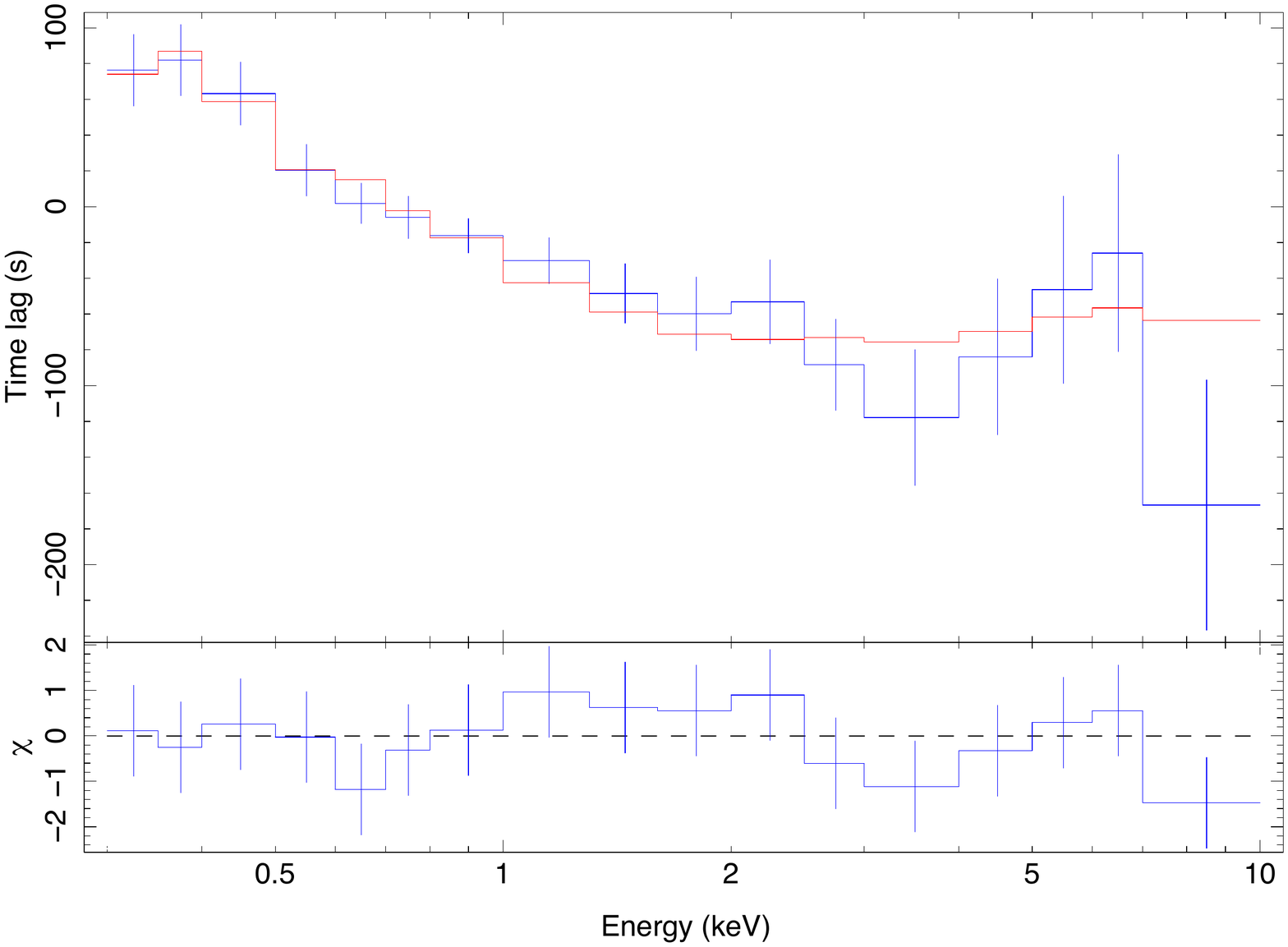}
}
\caption{Data and residuals from simultaneously fitting the time-averaged and lag-energy spectra of Mrk~335 (top panels), IRAS~13224-3809 (middle panels) and Ark~564 (bottom panels). The data and the model are shown in blue and red, respectively. The individual spectral-model component, modified by the warm absorbers where necessary, are shown in orange. For all AGN these components are the continuum, blurred reflection due to reverberation, {\sc reflionx} neutral reflection that gives the narrow line at 6.4~keV and a blackbody component that contributes significant flux in the soft band. Only Mrk~335 has a {\sc mekal} thermal component that gives the narrow emission line at $\approx 6.9$~keV. Ark~564 also has an additional broad Gaussian line at the centroid energy $\approx 0.4$~keV. We do not show the individual timing-model components since adequate fits are obtained without the power-law lags. See text for more detail. }
\label{fit_results}
\end{figure*}

Note that we are looking for a plausible model that provide good, simultaneous fits for both spectroscopic and timing data, not the best fits for each independently. Also the combined models for the full-reflection and ionized disc are complex, as discussed in \cite{Chainakun2015}, so we choose to step through all grid cells rather than interpolate between values which might find many local minima in a complex $\chi^2$ space. The fitting results of three AGN are presented in Fig.~\ref{fit_results}. However, as we have different model components between these AGN, we list in Table~\ref{tab:fit_para} only the key parameters constrained by the {\sc revb} model that relates to the inner-disc X-ray reverberation. The parameters of additional model components are specified in the text.

In general, we find that the {\sc revb} model as shown in Table~\ref{tab:fit_para} provides a plausible explanation for both time-averaged and time-lag properties of the AGN under the lamp-post assumption. This will be discussed in the next section. All AGN investigated here require a warm absorber, distant reflection and thermal blackbody-like emission at a temperature of $\approx 0.1$~keV. The distant reflection produces a small narrow Fe K line at $\approx 6.4$~keV. Mrk 335 requires an additional thermal emission component from distant ionised gas whose temperature is $\approx 9.8$~keV, producing an emission at $\approx 6.9$~keV. IRAS~13224-3809, on the other hand, has a photon index for distant neutral reflection of $\approx 3.3$ which is significantly larger than the value found for the {\sc revb} model. We also find that for Ark 564, both spectrum and time lags require an additional broad Gaussian line at $\approx 0.4$ keV (the centroid energy and width are tied between datasets). It can represent the component of reprocessed black body emission that is varying coherently with the reflection, or represent the blurred reflection flux in case the ionized gas producing $\approx 0.3-0.5$~keV emission lines has the atomic abundance more than one solar abundance. Without this component the model of Ark 564 can fit only between 0.7--10 keV. Last, but not least, it turns out that the power-law lags have only a small contribution compared to the reverberation lags. Their slopes are small, $-0.01, -0.2$ and 0 for Mrk 335, IRAS 13224-380 and Ark 546, respectively. Adequate fits therefore can be achieved without this component.	

The reported errors correspond to 90\% confidence intervals around the best-fitting parameters. Note that the full calculation is computationally intensive so we estimate the errors of each parameter by stepping through the grid cells of that parameter when other parameters of {\sc revb} model are fixed at their best-fitting values. Those from other models except {\sc revb} model are allowed to be free parameters. Linear interpolation is used if necessary. In the future errors will be calculated in which all free parameters are allowed to vary, and a finer grid of parameters will be computed around the best-fit values.

\begin{table}
\begin{tabular}{lp{1.0cm}p{2.0cm}l}
\hline
\multirow{2}{*}{{\sc revb} parameter} & \multirow{2}{*}{Mrk 335} &  \multirow{2}{*}{IRAS 13224-3809} & \multirow{2}{*}{Ark 564}\\  
 \\ \hline \vspace{0.1cm}
$h (r_g)$ & $3.0^{+0.1}_{-0.3}$ & $2.0^{+0.1}_{-0.1}$ & $5.0^{+0.6}_{-0.1}$ \\ \vspace{0.1cm}
$i$ ($^{\circ}$) & $45^{+1}_{-5}$ & $60^{+1}_{-1}$ & $45^{+1}_{-1}$ \\ \vspace{0.1cm}
$\Gamma$  & $2.3^{+0.1}_{-0.1}$ & $2.4^{+0.1}_{-0.1}$ & $2.6^{+0.1}_{-0.1}$ \\ \vspace{0.1cm}
$A$  & $2^{+3}_{-1}$ & $15^{+5}_{-5}$ & $1^{+1}_{-0.5}$\\ \vspace{0.1cm}
log $\xi_\text{ms} (\text{ erg cm s}^{-1})$ &  $4.7^{+0.1}_{-0.1}$ & $3.5^{+0.1}_{-0.1}$ & $3.8^{+0.2}_{-0.1}$ \\  \vspace{0.1cm}
$p$ &  $1.8^{+0.1}_{-0.1}$ & $2.3^{+0.2}_{-0.1}$ & $1.9^{+0.2}_{-0.1}$\\ \vspace{0.1cm}
$R_S$ & $0.9^{+0.1}_{-0.1}$ & $1.7^{+0.1}_{-0.1}$ & $0.2^{+0.1}_{-0.1}$\\ \vspace{0.1cm}
log $M (M_\odot)$ &  $7.13^{+0.22}_{-0.18}$ & $6.83^{+0.15}_{-0.04}$ & $6.60^{+0.17}_{-0.25}$\\  

\hline
$\chi^{2} / \text{d.o.f.}$ (mean data) & 1.16 & 1.17 & 1.37\\
$\chi^{2} / \text{d.o.f.}$ (lag data) &  1.49 & 1.98  & 1.03\\
$\chi^{2} / \text{d.o.f.}$ (combined) & 1.17 & 1.18 & 1.36\\
\hline
\end{tabular}
\caption{The best-fitting {\sc revb} parameters for simultaneous fits the time-averaged and lag-energy spectra of three AGN. The model parameters and the parameter values of Mrk 335, IRAS 13224-3809 and Ark 564 are listed in Columns 1, 2, 3 and 4, respectively. The errors correspond to 90\% confidence intervals around the best-fitting parameters estimated by linear interpolation between the model grid-cells. If the changes of $\chi^{2}$ are too large between adjacent grid-cells, the error estimate is given to be the grid spacing for that parameter. From the combined fits, the global $\chi^{2} / \text{d.o.f.}$ is provided together with the partial $\chi^{2} / \text{d.o.f.}$ for the time-averaged and lag-energy spectra only. Other parameters that are not related to the inner-disc reverberation are specified in the text. \label{tab:fit_para}}
\end{table}
   
\section{Discussion}

The three AGN investigated here show the traditional features of reverberation lags, the RDC ($\approx$ 0.3--1 keV or 5--7 keV) lagging behind the PLC ($\approx$ 2--4 keV). We perform simultaneous fitting of the lag-energy and time-averaged spectra. We find that the model can provide a good explanation of the data even though we have to add non-variable {\sc blackbody} and {\sc xstar} components. This suggests the thermal, possibly disc-like, emission in AGN and the warm absorber do not vary on the short timescales of inner disc reflection. We find that Mrk~335 has a source height of $\approx 3r_g$, which is low but consistent with the results of lag-frequency modelling \citep{Emmanoulopoulos2014} and combined spectral-timing \citep{Chainakun2015}. The central black hole mass is $\approx 1.35 \times 10^{7} M_\odot$, in agreement with the mass that follows the scaling relations \citep{Demarco2013}. Other parameters found here are quite comparable to those reported by \cite{Chainakun2015}, except that the iron abundance and the ionization at the innermost part are higher. The distant reflection from neutral material (e.g., cold torus) and thermal plasma at temperature $\approx 9.8$~keV are required to produce the X-ray emission at $\approx 6.4$ and 6.9~keV, respectively. The result are consistent with the framework in which the reflection from both the inner disc and distant material such as the cold torus and the ionized gas filling that torus as presented by, e.g., \cite{Oneill2007}. The reflection from such distant material does not vary at the reverberation time-scales, so including it in the spectral model does not affect the timing model.

The same spectrum of IRAS 13224-3809 has been studied by \cite{Fabian2013} who suggested a patchy disc model producing two different reflection spectra from different ionized materials, the narrow Gaussian line ($\approx6.4$ keV) and the thermal disc emission. Here we present an alternative model based on a simple ionized disc which can also fit the timing data of IRAS 13224-3809. Our model consists of one blurred reflection spectrum ({\sc revb}), one unblurred distant-reflection spectrum (unblurred-{\sc reflionx}) and one non-variable disc thermal component ({\sc blackbody}). We find this AGN has very strong overabundance of iron, $A=15$, consistent with \cite{Fabian2013}. The X-ray source is very close to the black hole, $h \approx 2r_{g}$, which is expected as the observation was made during its low flux state. However, the photon index significantly differs between blurred ($\Gamma = 2.4$) and unblurred components ($\Gamma = 3.4$), suggesting that the distant neutral reflection contributes significantly in the soft band. We interpret this as the distant material also possessing a high abundance of other atoms, in addition to iron, that are responsible for the soft band emission. Allowing the abundances of such atoms to vary in our fits is computationally too expensive so all abundances except that of iron are fixed at the solar value. Furthermore, the corresponding lag-energy model produces a good fit without contribution from the additional unblurred component. Our lag-energy fitting therefore suggests that the soft excess in IRAS 13224-3809 has a distant reflection origin. This may explain why the fractional RMS variability spectrum \citep{Fabian2013} shows that the soft excess of IRAS 13224-3809 does not vary as much as the 2--10~keV band. Such behaviour of the RMS spectrum could be understood in terms of either light bending effects or a distant reflection origin of the soft-excess. 

Previous studies using optical observations \citep[e.g.,][]{Boller1993,Kaspi2000} implied the central mass of IRAS 13224-3809 to be $M \approx10^{7} M_{\odot}$. 
\cite{Kara2013a} used a mass estimate of $\approx 5.8 \times 10^6 M_{\odot}$ reported by \cite{Zhou2005} to show that it is in a linear relationship with the amplitude of the Fe K lag along with those of other six AGN. Recently, \cite{Chiang2015} performed a series of spectral fitting and deduced a black hole mass to be $M \approx 3.5 \times 10^6 M_{\odot}$ from the blackbody emission constrained by {\sc kerrbb} \citep{Li2005}, a multi-temperature blackbody model for a relativistic thin accretion disk around a Kerr black hole. Using the combined spectral and timing analysis, our mass is $M \approx 6.8 \times 10^{6} M_{\odot}$, very similar to what was previously found.

On the other hand, we find that Ark 564 requires both inner-disc and distant reflection components, but the combined model successfully fits the data only from 0.7--10~keV. While the soft excess of IRAS 13224-3809 is dominated by the narrow components, the soft excess of Ark 564 is dominated by the broad components. Below 0.7 keV both spectral and timing models under-predict the data of Ark 564 and an additional broad Gaussian component is added to obtain the fits shown in Fig.~\ref{fit_results}. Our model of Ark 564 also requires a warm absorber that covers only the distant reflection spectrum which is a contrived but possible geometry as reported by \cite{Giustini2015}. The narrow components contribute only to the spectral model while the broad components contribute to both spectral and timing models. The additional line added to the soft excess band of Ark 564 is broad so it has to originate close to the black hole and is varying at the reverberation time-scales. We therefore tie the centroid-energy and width of the Gaussian line across the datasets. This can be understood if the atomic abundance of ionized gas at the innermost region responsible for the X-ray emission at $\approx$ 0.3--0.4~keV exceeds the solar abundance. Optional interpretation of this broad Gaussian line is that it is the reprocessed black body emission which is varying coherently with the reflection component. The high photon index, $\Gamma=2.6$, is consistent with \cite{Giustini2015}. The central mass of Ark 564 is quite low, $M\approx 3.9 \times 10^{6} M_{\odot}$, comparable to the value found by \cite{Botte2004}. 

Most importantly, the 3 keV dip and the $>7$~keV drop in the lag-energy spectra of these AGN are very difficult to produce. The $\approx 7-10$~keV band leading the reference band has not been well addressed before in the lag-frequency analysis because the 7--10~keV band is not usually selected as the PLC band. Recently, \cite{Wilkins2016} have shown that the 3~keV dip is successfully produced by luminosity fluctuations that slowly propagate upward along a vertically extended corona. In that case the lower part of the corona produce the lead of $\approx 3$~keV band that responds first at the innermost part before the $> 3$~keV bands respond later from the outer regions. The fluctuations then modulate the upper corona and produce the delayed continuum with less gravitational focusing on the disc so that the 1--2~keV continuum dominated band lags behind the 3 keV band due to the propagation time delays. They suggest that the 3~keV dip might not be produced by a standard lamp-post model as the detection of the reflection and its primary continuum from a single point source should always result in the 3~keV band lagging behind the 1--2~keV band which is the continuum dominated. However, our models suggest that the 3~keV dip and the 7--10~keV drop are possibly produced if the effects of ionization gradients in the disc are taken into account. This requires either the source height to be $\gtrsim 5r_g$ or the disc is highly ionized at the innermost part and is colder further out. The emergent spectra from different parts of the disc are different and hence the response function of a specific band depends on both the redshifts and the intrinsic spectral shapes. The lags, on the other hand, depend on the response functions and the contribution between the continuum and reflection flux (i.e., the RRF). If the innermost regions are highly ionized, the reverberation signatures produced through reflection from these regions (e.g., the 3 keV lags due to the redshifted iron K$\alpha$ photons) are less observable. The prominent emission lines can be seen only from colder reflection off the outer part. These lines are likely to be redshifted or blueshifted to other energies rather than 3~keV giving the RRF in 3~keV band smaller than those of the adjacent bands, and, as a result, the 3~keV dip appears in lag-energy profile. Besides, this is the reason why the 3~keV and $> 7$~keV bands of IRAS 13224-3809 are not fit very well. The model of IRAS 13224-3809 has moderate $\xi_\text{ms}$ and low source height meaning that the 3~keV dip and the 7--10~keV drop cannot be produced.

\begin{figure}
\vspace{-0.7cm}
\centerline{
\includegraphics*[width=0.55\textwidth]{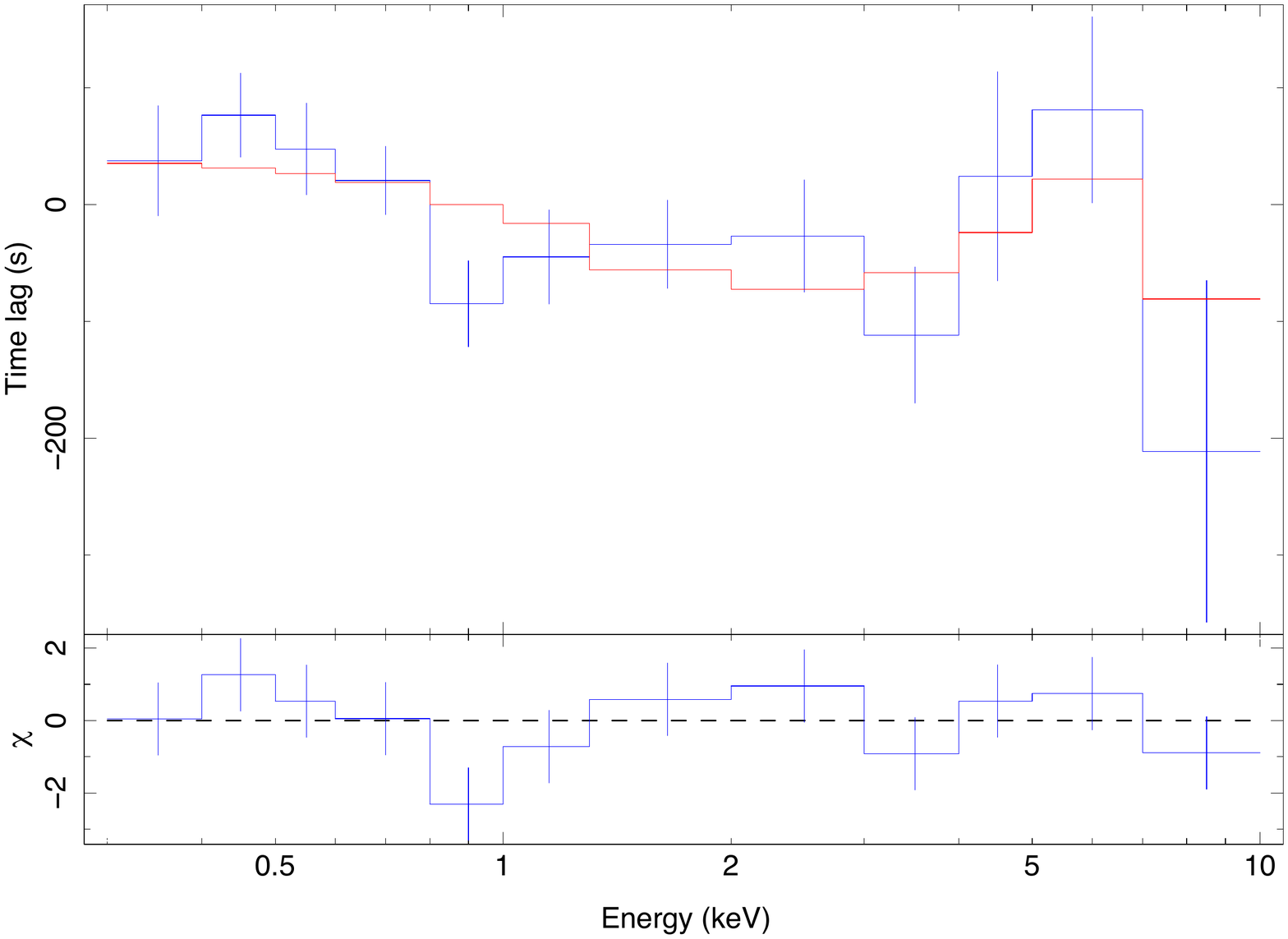}
}
\vspace{-0.8cm}
\centerline{
\includegraphics*[width=0.50\textwidth]{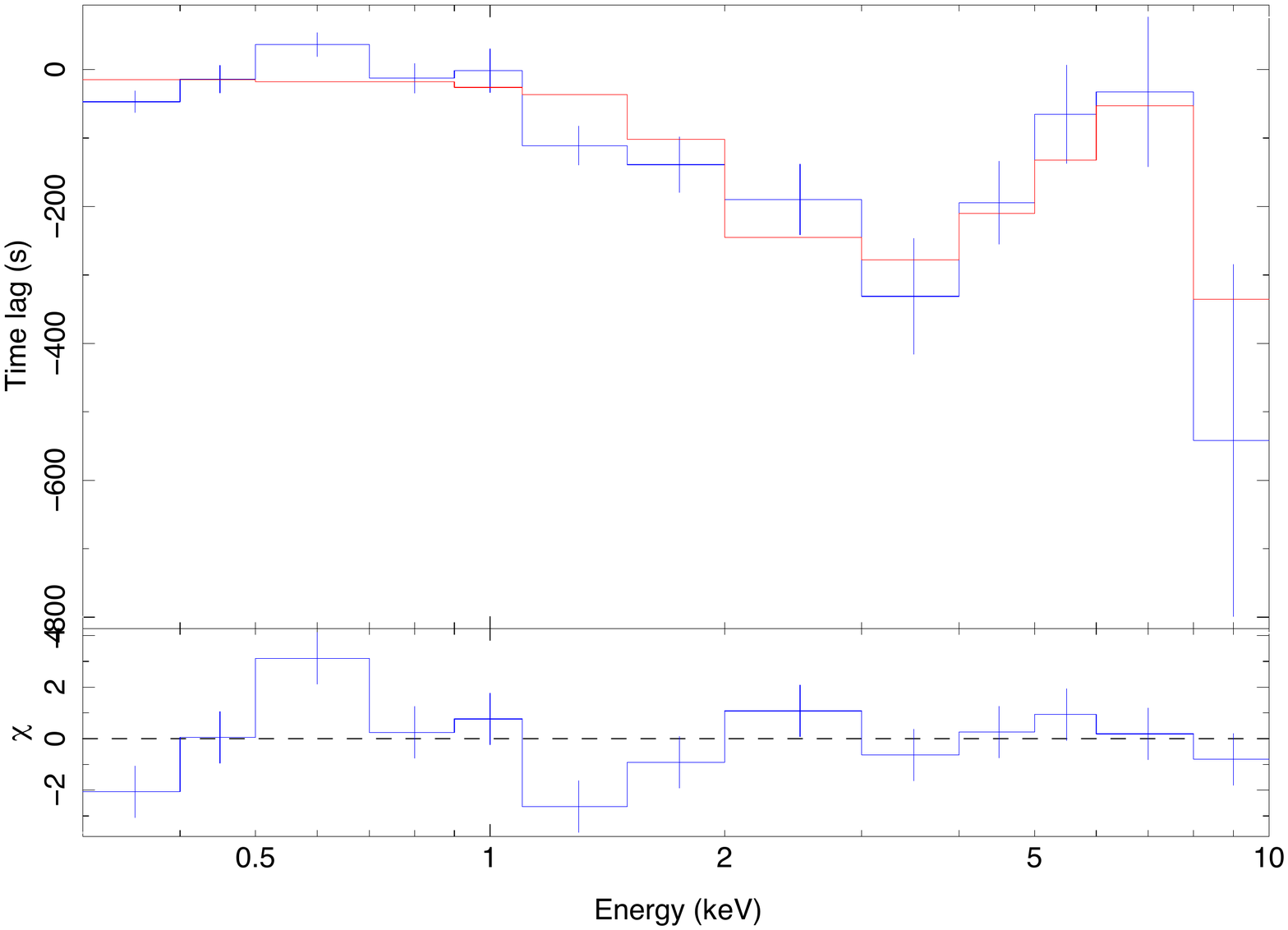}
}
\vspace{-0.8cm}
\centerline{
\includegraphics*[width=0.55\textwidth]{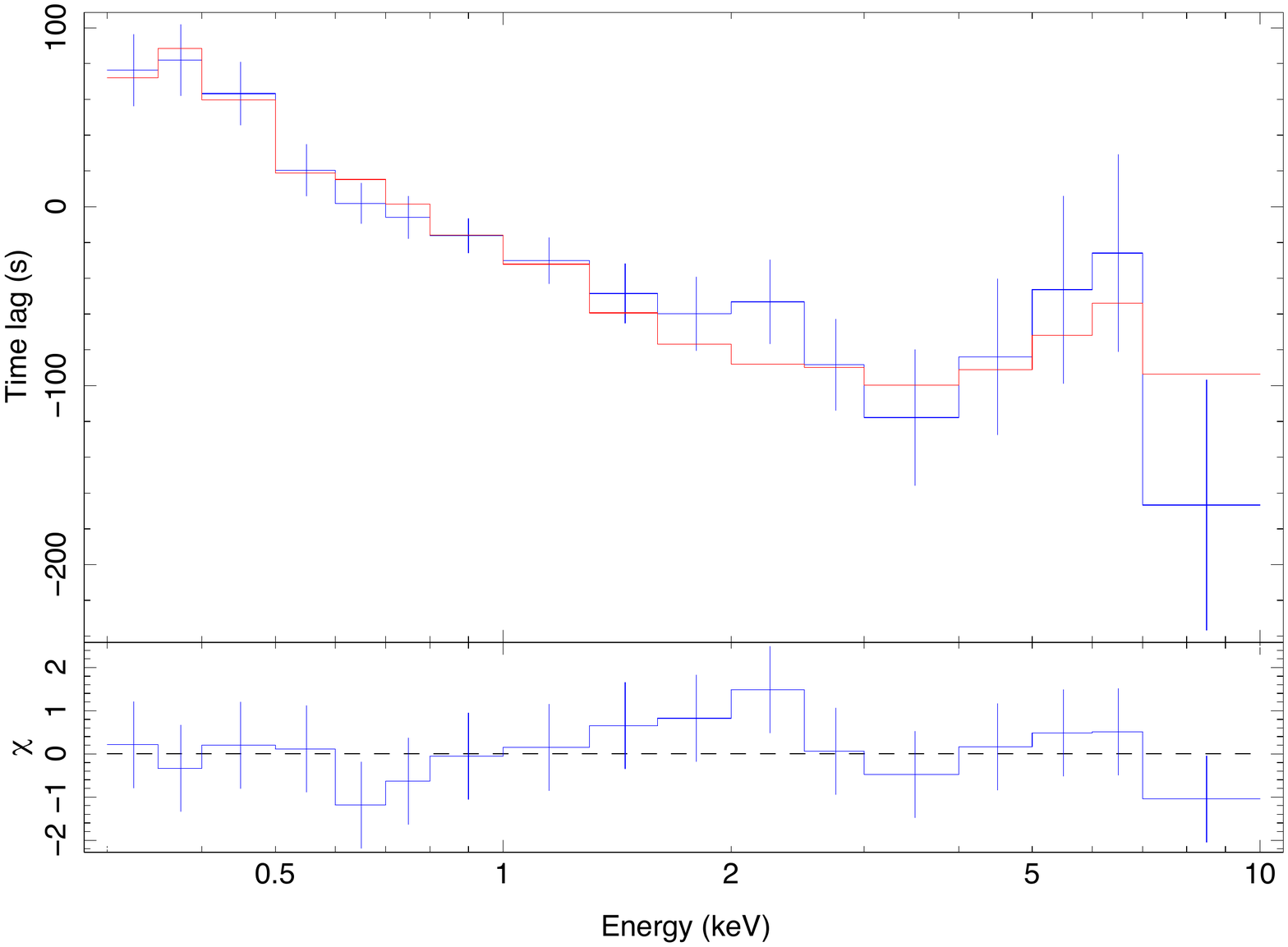}
}
\caption{Data and residuals from fitting the {\sc revb} model to only the lag-energy spectrum of Mrk~335 (top panel), IRAS 13224-3809 (middle panel) and Ark 564 (bottom panel). The fitting-model expression of each AGN is similar to that is shown in Table~\ref{tab:model_expression}. }
\label{fit_lags}
\end{figure}

Fig.~\ref{fit_lags} shows the results when the lag-energy spectrum of each AGN is fitted alone. The slopes of power-law lags are $-1.1$ for Mrk~335 and 0 for both IRAS~13224-3809 and Ark~564. Some of the fitting parameters are different to those found when the spectral and timing data are combined. Mrk 335 ($\chi^{2} / \text{d.o.f.}=1.27$) has the X-ray source at $4r_{g}$, a more highly ionized disc, $\xi_\text{ms} \approx 10^{5} \text{ erg cm s}^{-1}$, and a bigger black hole mass, $M \approx 1.78 \times 10^{7} M_{\odot}$. For IRAS~13224-3809 ($\chi^{2} / \text{d.o.f.}=1.65$) the model, without power-law lags, can capture the dips at 3--4 keV and 7--10 keV very well and requires $h=5r_{g}$ and $\xi_\text{ms} \approx 3.2 \times 10^{4} \text{ erg cm s}^{-1}$, proving the concept that the large source height and the ionization gradient where the disc is highly ionized at the innermost part naturally produce the drop of the lags at 3~keV and $> 7$~keV. The lower iron abundance, $A=5$, is found but to enhance the strong Fe K lags the model requires a larger black hole mass, $M \approx 8.3 \times 10^6 M_\odot$. In case of Ark 564 ($\chi^{2} / \text{d.o.f.}=0.91$), the model still requires the broad Gaussian line at $\approx 0.4$~keV and fitting the lag-energy spectrum alone places the source at $h=8r_{g}$. The ionization is higher, $\xi_\text{ms} \approx 1.5 \times  10^{4} \text{ erg cm s}^{-1}$, while the constrained mass does not change, $M \approx 3.9 \times 10^{6} M_{\odot}$. On the other hand, if the mean spectrum is fitted alone (using the spectral-model expression shown in Table~\ref{tab:model_expression}), we found the fits are not much improved. While the inclination and photon index remain the same as those fits with the lag spectrum (Table~\ref{tab:fit_para}), the inner disc seems to have lower ionization. Our results suggest that the lag-energy spectra supports a higher source height and higher ionisation, while the mean spectra supports lower source height and lower ionisation. This, therefore, shows a problem in the fitting model when we combine spectral and timing data.

\begin{figure}
\centering  
\vspace{-0.7cm}
\includegraphics*[width=95mm]{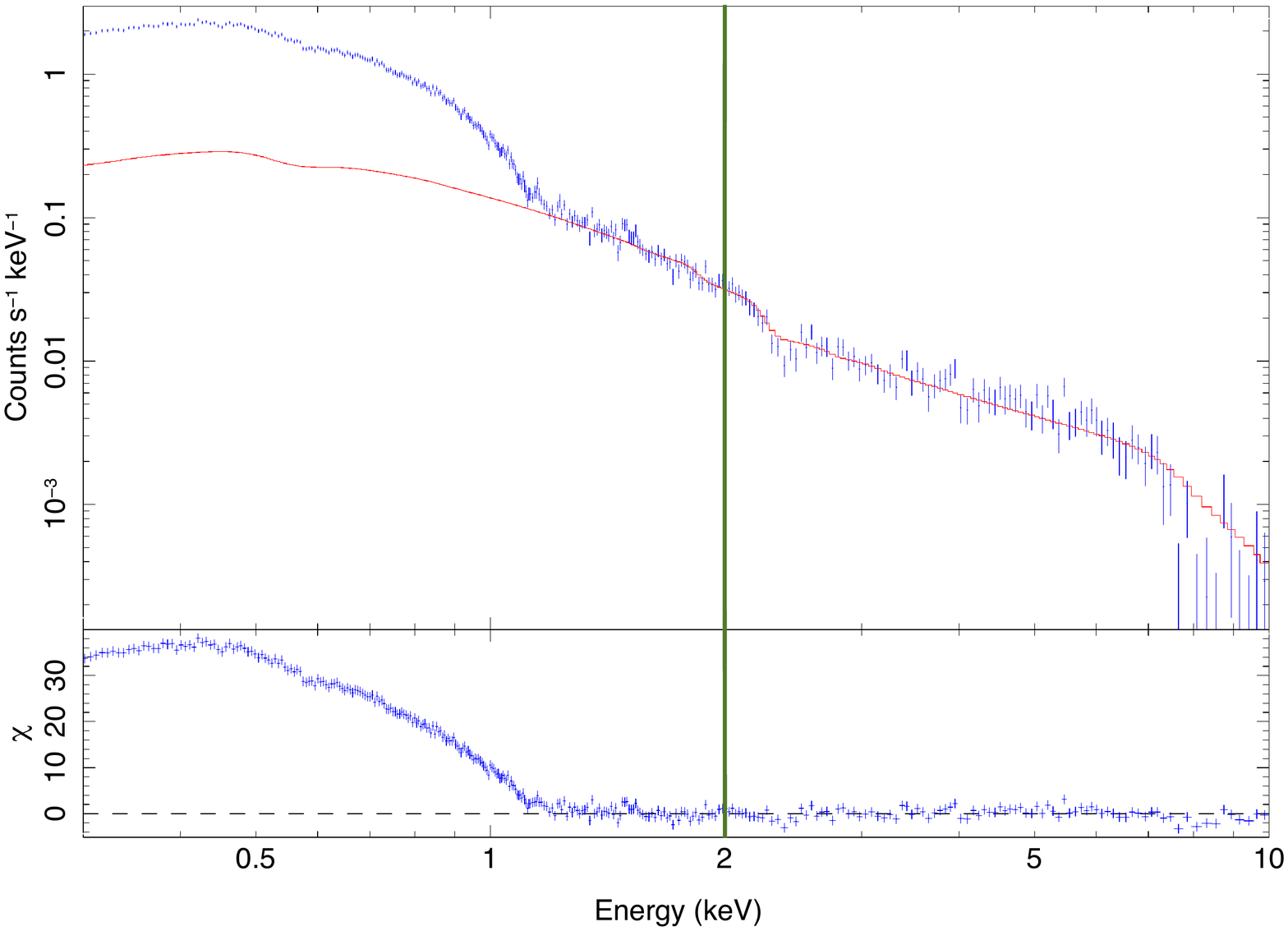}
\vspace{-0.8cm}
\caption{Data and residuals when the 2--10~ keV time-averaged spectrum of IRAS~13224-3809 is reproduced with the best fit parameters of $\textsc{revb}$ model obtained from fitting the lag-energy spectrum alone (as shown in middle panel of Fig.~\ref{fit_lags}). We exclude all other spectral components and extrapolate this fit to lower energies so that the residuals reveal all components which are not variable on reverberation timescales. The extrapolation and the fitting bands are separated by the vertical green line.
\label{spec_iras_interp}}
\end{figure}

\begin{figure}
\centering  
\vspace{-0.7cm}
\includegraphics*[width=95mm]{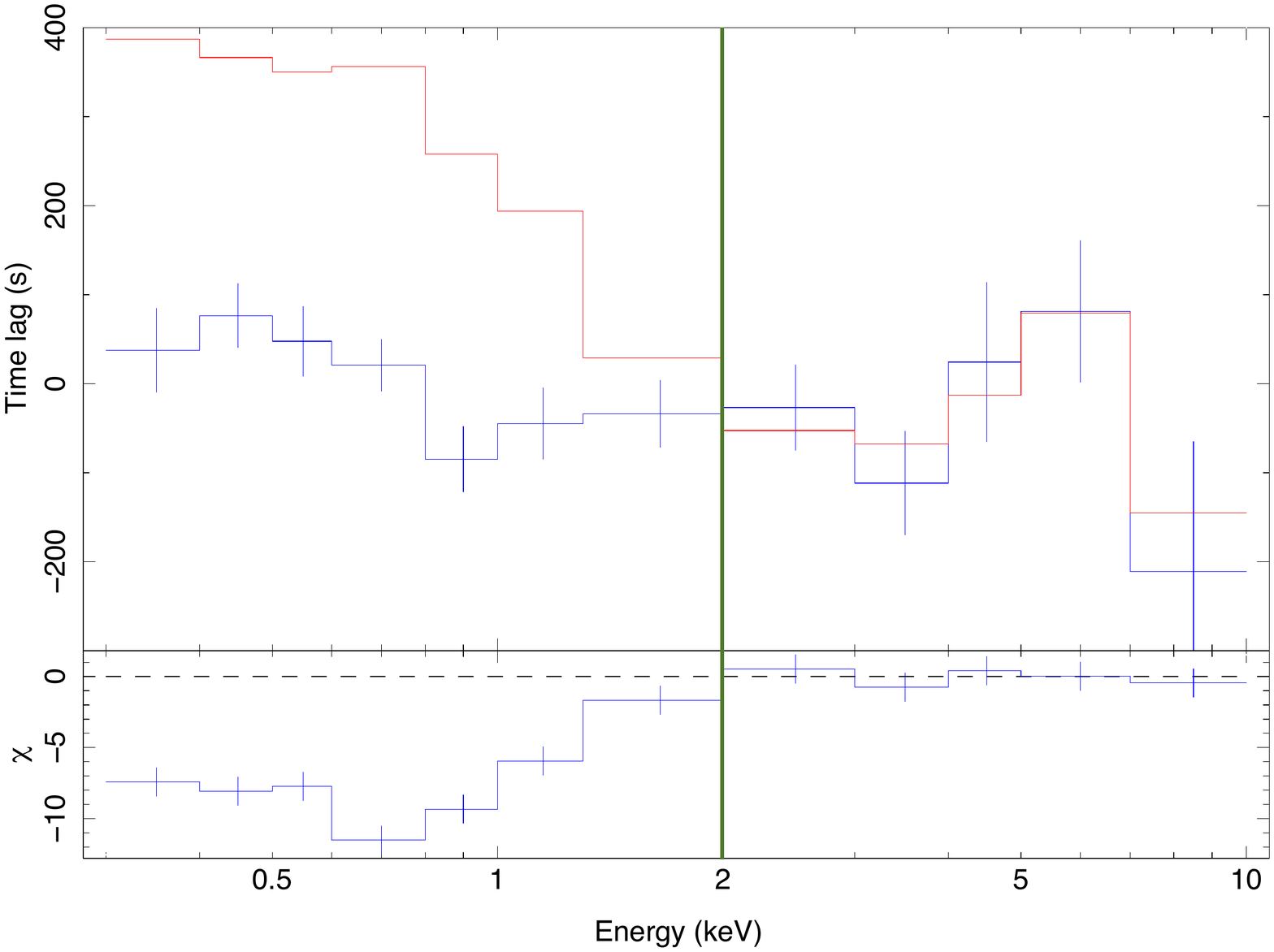}
\vspace{-0.8cm}
\caption{Data and residuals from fitting the {\sc revb} model to the 2--10~keV time lags of Mrk 335. The sharp drop of lags at $> 7 $~keV is well produced, but extrapolating this fit to lower energies shows that this model over-predicts lags in the 0.3--2~keV band. The extrapolation and the fitting bands are separated by the vertical green line.
\label{h_lag_mrk335}}
\end{figure}

In principle, the lag-energy profile should follow the shape of the mean spectrum if all spectral components are variable on the short timescales of the inner-disc reflection. Fig.~\ref{spec_iras_interp} shows an example in the case of IRAS~13224-3809 when its time-averaged spectrum is reproduced with the best fit parameters from the lag-energy spectrum alone. We use only the $\textsc{revb}$ parameters from the fits in middle panel of Fig.~\ref{fit_lags} and exclude all other spectral components to reproduce the mean spectrum in 2--10~keV band. Since the lags are fitted from 0.3--10~keV, extrapolating the spectral fits to 0.3~keV in this case suggests the presence of the non-variable (or at least uncorrelated) components in the soft band that do not contribute to the time lags. We do not quantify these components but it is interesting to note that other possible models could exist. Moreover, if we fit only the lags and only from 2--10~keV, the model provides a good fit to the dip at 3~keV, the strong Fe K lags and the sharp drop of lags at $> 7$~keV. Fig.~\ref{h_lag_mrk335} represents an example in the case of Mrk~335 in which the soft band fit, from 0.3--2~keV, has been extrapolated from the hard band fit. The over-prediction of the soft excess band is clearly seen suggesting an alternative model that the lags are diluted by the complex soft excess components, or that the corona is extended and the entire corona is not varying coherently, but we have not investigated these possibilities in this paper. How the lags change requires a well defined timescales for the variability of those additional components (e.g., blackbody emission) which is beyond the scope of this paper. As pointed out by \cite{Chainakun2015}, the soft excess lags are ambiguous and possibly model-dependent which make constraining this band very difficult. We therefore focus on the simpler model that can explain both spectroscopic and timing properties of these AGN (Fig.~\ref{fit_results}).

Although the observed lags are produced at frequencies dominated by reverberation, we should expect a small contamination from the power-law lags (e.g., propagation lags) whose amplitude increases with energy. This was suggested by \cite{Arevalo2006} that the lag amplitude increases with the difference between the emissivity indices of the light curves. However we find no clear evidence of the power-law lags in the lag-energy spectra. The error bars in the Fe~K band are large which make distinguishing the combined components (e.g., reverberation and power-law lags) in this band very difficult. The power-law lags constrained here have very small normalization and gradient (gradient $= 0$ in case of Ark~564) so they do not play a major role in fitting the data. Also fitting the Compton hump lags using \emph{NuSTAR} data might provide more clues as to the nature of the power-law and reverberation lags which will be investigated in a future work. Furthermore, \cite{Epitropakis2016} suggests that there is possibly a bias in time-lag estimates using standard Fourier techniques. Epitropakis et al. (2016b, in prep) fit the frequency-dependent Fe~K lags of seven AGN and find that the potential to constrain key parameters using the lag-frequency spectra alone should be questioned. We plan to apply their new time-lag estimation techniques to investigate lag-energy spectra in the future. Ultimately, the time-averaged and lag-energy spectra are what observers have been presenting in the literature, but fitting the full cross-spectrum instead might be preferable, although this needs more investigation to identify which statistical approach is optimal. Furthermore, in our present fits it is not clear what the relative weighting between the time-averaged and lag-energy spectra should be (e.g., compare the fits in Fig.~\ref{fit_results} and Fig.~\ref{fit_lags}).

Last, but not least, there are several ways that our model can be improved, even though all require very intensive computations. The fits will be improved if we interpolate over key parameters such as the photon index, $\Gamma$, and the iron abundance, $A$, that significantly affect the shape of the spectrum. Allowing more parameters (e.g., the spin parameter, $a$, and the size of an accretion disc) to vary is another option but it can lead to many degeneracies of the model. We note that here the reflected response fraction is assumed to be an averaged function of only the energy bin, $R=R(E)$, but in principle it can change over time with the variations of the X-ray continuum. Assuming $R=R(E,t)$ is therefore a realistic approach and is worth investigating in the future. Finally, investigating extended rather than point source corona will be important.

\section{Conclusion}

Fitting the combined spectral-timing data provides a self-consistent explanation of the rapidly variable X-ray phenomena in AGN. Of three AGN we investigate, our model supports the inner-disc reflection scheme in which the observed lags are results of the delays between the continuum and the blurred reflection spectra. Additional neutral reflection from distant material is required but it makes no contribution to the lags since its variation is on much longer timescales than that of inner-disc reverberation, as also suggested by \cite{Chainakun2015}. A blackbody-like emission component that dominates the soft-excess band is also required but how it varies is still unclear for AGN. We find that good fits are obtained even though we assume a non-variable blackbody component. Another component not included in the current model is the power-law lags. However, we find the slopes of power-law lags are small for three AGN we investigated, suggesting that they do not play an important role at reverberation timescales.

For Mrk 335, the model prefers the framework of \cite{Oneill2007} where the spectrum is a combination of reflection from the inner accretion disc, reflection from distant neutral material and thermal emission from an ionised gas. The distant material in this case can be a cold torus filled by an ionzied plasma. This is consistent with the spectral and lag-frequency analysis performed by \cite{Chainakun2015}. For IRAS~13224-3809, our model differs from the patchy disc model suggested by \cite{Fabian2013}. We find that a simple ionized disc can still provide good fits to both spectrum and time-lags. The model also supports a distant reflection origin for the soft excess of IRAS~13224-3809. The data of Ark~564, however, cannot be explained by a simple model. An additional broad Gaussian is needed to fit both spectral and timing profiles in the soft excess band. Its origin is therefore from the inner disc reflection where the relativistic effects broaden the narrow features. This can be interpreted as the abundances of atoms responsible for the emission at soft energy band are more than the solar value we have used in the {\sc reflionx} model.

We emphasize that it is necessary to combine the spectral and timing data in order to obtain a feasible model that provides self-consistent fits, which is achievable by our model. Although alternative models could exist, the reduced $\chi^2$ values for all sources are reasonable, leading to the conclusion that the lamp-post model is still a reasonable framework to understand their X-ray properties close to the black hole. Most of the characteristic features seen in the time-averaged and lag-energy spectra can be reproduced by the reflection model. However, the ionized disc illuminated by a single point source can provide the clear 3~keV dip and $ > 7$~keV drop only for a limited range of scenarios in which the source height is large ($\gtrsim 5r_g$) and the disc is highly ionized around the centre but colder further out. The model can also be straightforwardly extended to investigate Compton hump reverberation lags obtained by \emph{NuSTAR}, and to model extended sources, both of which are planned for the future.

\section*{Acknowledgements}
This work was carried out using the computational facilities of the Advanced Computing Research Centre, University of Bristol. PC thanks the University of Bristol for a Postgraduate Research Scholarship. AJY would like to thank the organisers and participants of the Lorentz Center Workshop on ``The X-ray Spectral-Timing Revolution'' for an interesting meeting and helpful discussions. EK thanks the Hubble Fellowship Program for support under grant number HST-HF2-51360.001-A from the Space Telescope Science Institute, which is operated by the Association of Universities for Research in Astronomy, Incorporated, under NASA contract NAS5-26555. We thank the anonymous referee for useful comments which have improved the paper.




\bibliographystyle{mnras}

\bibliography{spec_reverb}








\bsp	
\label{lastpage}
\end{document}